\documentclass[fleqn,usenatbib]{mnras}

\usepackage[T1]{fontenc}

\DeclareRobustCommand{\VAN}[3]{#2}
\let\VANthebibliography\thebibliography
\def\thebibliography{\DeclareRobustCommand{\VAN}[3]{##3}\VANthebibliography}

\usepackage{graphicx}	
\usepackage{amsmath}	
\usepackage{amssymb}	

\usepackage{newtxtext,newtxmath}
\usepackage{bm}
\usepackage{epstopdf}
\usepackage{verbatim}
\usepackage{natbib}
\usepackage{hyperref}
\usepackage{float}
\usepackage{color}
\usepackage{footnote}
\usepackage{threeparttable}%

\title[CAgNVAS I: new generation DIFMAP]{\textit{A new generation DIFMAP for Modelfitting Interferometric Data and Estimating Variances, Biases and Correlations}}

\author[A. Roychowdhury et al.]{
Agniva Roychowdhury,$^{1}$\thanks{E-mail: agniva.physics@gmail.com}
and Eileen T. Meyer$^{1}$
\\
$^{1}$Department of Physics, University of Maryland Baltimore County, 1000, Hilltop Circle, Baltimore MD 21250, USA\\
}

\date{Accepted XXX. Received YYY; in original form ZZZ}

\pubyear{2023}

\begin{document}
\label{firstpage}
\pagerange{\pageref{firstpage}--\pageref{lastpage}}
\maketitle

\begin{abstract}
We present the program `Catalogue of proper motions in extragalactic jets from Active galactic Nuclei with Very large Array Studies' or CAgNVAS, with the objective of using archival and new VLA observations to measure proper motions of jet components beyond hundred parsecs. This objective requires extremely high accuracy in component localization. Interferometric datasets are noisy and often lack optimal coverage of the visibility plane, making interpretation of subtleties in deconvolved imaging inaccurate. Fitting models to complex visibilities, rather than working in the imaging plane, is generally preferred as a solution when one needs the most accurate description of the true source structure.  In this paper, we present a new generation version of $\texttt{DIFMAP}$ (\texttt{ngDIFMAP}) to model and fit interferometric closure quantities developed for the CAgNVAS program. \texttt{ngDIFMAP} uses a global optimization algorithm based on simulated annealing, which results in more accurate parameter estimation especially when the number of parameters is high. Using this package we demonstrate the ramifications of amplitude and phase errors, as well as loss of $u-v$ coverage, on parameters estimated from visibility data.  The package can be used to accurately predict variance, bias, and correlations between parameters. Our results demonstrate the limits on information recovery from noisy interferometric data, with a particular focus on the accurate reporting of errors on measured quantities.

\end{abstract}




\section{Introduction}

A radio interferometer is a collection of two or more radio antennas where each baseline samples a complex visibility, approximately described as the Fourier transform of the brightness distribution of the observed target. Therefore, an accordingly constructed baseline allows a radio array to attain an "effective" diameter much larger than a single antenna, and thereby angular resolutions well beyond the diffraction limit. \textit{Directly} model-fitting complex visibilities reduces a number of biases and errors that may otherwise be amplified by image plane analysis \citep{pearson99}. Visibility fitting is particularly useful when very precise determination of source parameters is required and/or when the Fourier space coverage (also known as $u-v$ coverage) is non-optimal. For example, the accuracy of the image of the black hole in M87, observed by the Event Horizon Telescope using almost Earth-sized baselines, was verified by robust $u-v$ plane analysis \citep{eht19}. Additionally, "photon rings" inside the unresolved black hole shadow can be identified through specific predictions of model visibilities in the longest baselines in the future \citep{johnson20}, enabling possible determination of black hole mass and spin. 

In roughly 5\% of active galactic nuclei, the region close to this black hole launches bipolar collimated jets of plasma that extend through 10 orders of magnitude in gravitational radius. Despite numerous multi-wavelength studies of these jets over decades, present studies are generally plagued by large uncertainties and bias in important measures such as the energy content of the jet, effects of the jet viewing angle and the physical nature of the plasma when we only observe the emitted radiation.

Studies of proper motions of "knots", or bright compact components at different positions in these jets can, in principle, be used to produce model-independent constraints on the Lorentz factor ($\Gamma$) of the bulk motion, which can be used to provide robust answers to open questions mentioned above \citep[see e.g.][]{meyer16_3c273}. Proper motion studies of these jets exist in plenty at the parsec and sub-parsec scales, observed as a part of many Very Long Baseline Interferometry (VLBI) monitoring programs by different groups \citep[e.g.,][]{reid89,junor95,zensus95,tingay98,jorstad01,homan01,middel04,muller14_tanami,lister16,piner18,walker18}. The standard approach is to fit Fourier transforms of two dimensional Gaussian models of knots to the complex visibilities, and thereafter to track the central peak position through subsequent epochs. Combining proper motions with polarization and opening angle measurements, these studies have produced strong statistical constraints on the velocity, structure, magnetic field and dynamical evolution of a sample of approximately 500 $\lesssim$ sub-parsec- and parsec-scale jets. In contrast, studies of these jets on the larger $>100$ parsec scales are rare, with results for a few jets on the kiloparsec-scale \citep{reid89,biret95_hst,biret99,cheung07,kovalev07,ly2007,meyer13,asada14,meyer15, meyer16_3c273, snios19}. The construction of the Very Large Array (VLA) in the late 1970s allows for current time baselines $>40$ years for many kpc-scale radio jets, necessary to compensate for the much lower resolution compared to VLBI, in terms of measurement accuracy. Using the richness of the VLA archive as well as recently obtained observations, our goal is to create a large Catalogue of proper motions in Active galactic Nuclei using Very Large Array Studies (CAgNVAS), with a view to investigating physical properties of large-scale jets and their influence on the inter-galactic environment, both statistically and on a case-by-case basis. 

The general approach to measuring radio proper motions usually involves fitting of the complex visibilities with a tool called \texttt{DIFMAP} (Difference Mapping, \citealt{shep94}), using a set of simple model components by a local minimizer, or a Levenberg-Marquadt optimization algorithm. However, even though radio interferometric data can have multiple sources of error, the uncertainties on best-fit parameters are determined rather naively, simply from the curvature of the $\chi^2$ surface. Furthermore, the local minimizer is not ideal when high accuracies are needed. Starting from the original $\texttt{DIFMAP}$ package, we have created an enhanced "new generation" \texttt{DIFMAP} (\texttt{ngDIFMAP}) that contains a global optimizer and which fits interferometric closure quantities. The global optimizer allows more accurate determination of best-fit values of the source structure parameters. Usage of closure quantities partly removes the issue of dealing with antenna-based gain errors. Additionally, in order to determine the biases, variances and correlations between these parameters, a new functionality inside \texttt{ngDIFMAP} allows determination of approximate effects of errors in interferometric data on best-fit parameters. 

In this paper, we introduce $\texttt{ngDIFMAP}$ and discuss major sources of bias and related uncertainties in determining mean values of best-fit parameters in model-fitting radio interferometric datasets, with a special focus on jets from radio-loud active galactic nuclei (AGN). In Section \ref{sec:backg}, we describe the platform on which our tools have been built. In Section \ref{sec:code}, we describe the specifics of the new code. In Section \ref{sec:mc}, we observe the effect of amplitude and phase errors on the visibilities as well as the parameters using Monte Carlo simulations and the corresponding covariance matrix. In Section \ref{sec:disc}, we provide more insight into our results and cast light on possible future ventures. In Section \ref{sec:conc}, we conclude with a summary of our main results.

\section{Background}
\label{sec:backg}
\subsection{Preliminaries}

A radio interferometer is a collection of radio antennas, where each pair is known as a baseline. Each baseline measures the spatial coherence function of the electric field from an astronomical source through a correlator. The coherence function and the sky brightness distribution are related as  \citep{vancitt,thompson17}:

\begin{multline}
    V(u,v,w)=\int^{\infty}_{-\infty}\int^{\infty}_{-\infty}A(l,m)I(l,m)e^{-i[ul+vm+w(\sqrt{1-l^2-m^2}-1)]}\\ 
    \frac{dldm}{\sqrt{1-l^2-m^2}}
\label{eq:1}
\end{multline}

\vspace{0.5pc}
where $u$, $v$ and $w$ are the unit-less three-dimensional baseline coordinates scaled by the observing wavelength in an ideal monochromatic interferometer, $l$, $m$ and $n$ are the direction cosines of the source measured with respect to the $u$, $v$ and $w$ axes. 
$I(l,m)$ is the sky brightness distribution and $A(l,m)$ is the angle-dependent antenna response pattern, also known as the primary beam. $V(u,v,w)$ is a complex quantity and hence is also called the complex visibility. Inverting the above equation to produce $I(l,m)$ is the essence of imaging.

The type of datasets tested in this work relate to high-frequency ($\gtrsim 5$ GHz) Very Large Array (VLA) observations that generally have minimal background noise with majorly the target in the field of view ($\Theta\lesssim8'$) \citep{perley99b}. A compact source ($\sim$ few arcsec) implies that $l^2,m^2\ll1$ and Equation \ref{eq:1} reduces to a simple Fourier Transform relation. Every baseline samples one $(u,v)$ pair at a given instant of time. When observations are long enough, the rotation of the earth allows radio antennas to densely sample this $u-v$ plane, in a process known as earth-rotation aperture synthesis \citep{thompson17}. However, if the source is highly extended and the non-co-planarity due to long observations is taken into account, the current version of \texttt{DIFMAP} cannot process fitting models to the $u-v$ plane since it uses a direct Fourier Transform relation. This has not been considered in this work and will be discusssed in a future endeavour. In this work, we are only dealing with the observed visibility $V$ in the \textit{projected} $u-v$ plane ignoring imaging of any sort which, if required, would be given by a Fourier Transform of the visibilities.

For any pair of antennas represented by $i,j$, the actual observed visibility $\tilde{V}$ can be represented as:

\begin{equation}
\label{eq:vis}
    \tilde{V}_{ij}=H_{ij}G_{i}G^*_{j}[{V}_{ij}+\epsilon_{ij}]
\end{equation}

where $V$ is the true visibility and $\epsilon$ represents additive thermal noise. $G_i$ and $G_j$ are the corresponding final antenna gains, $H_{ij}(t)$ can represent a baseline-based correction, which are estimated \textit{after} flux and bandpass calibration. For this study we shall define a "real" dataset as one which has residual gain errors as well as imperfect $u-v$ coverage. For ease of modelling these errors, modifying the prescription of \cite{chael18}, we have more generally modelled $\tilde{V}$ with time-dependent errors as:

\begin{equation}
\label{eq:ge}
\tilde{V}_{ij}=[1+Y_{ij}(t)]e^{\bm{j}\delta\phi_{ij}}[{V}_{ij}+\epsilon_{ij}] 
\end{equation}

where $Y_{ij}$ is a random variable $\sim$ $\mathcal{N}(0,\sigma_g)$, which refers to a normal distribution with $\mu=0$ and $\sigma=\sigma_g$. $\sigma_g$ represents the level of gain error (1 for 100\%). $\delta\phi_{ij}$ is a random variable $\sim$ $\mathcal{N}(0,\sigma_p)$, with $\sigma_p$ referring to the phase error in radians. $Y_{ij}$ and $\delta\phi_{ij}$ are drawn for every ($u,v$ or $i,j$) point that is sampled. In general, it is impossible to have a single accurate prescription that would produce the imperfections of the radio array, consequent calibration and self-calibration in the antenna gains. Hence, to first approximation, we have used a Gaussian distribution, which will be found to mimic the scatter in real data adequately (Section \ref{sec:det_err}). 

\subsection{Parameter Variances and the Monte Carlo Method}


The "independence" of the real and imaginary visibilities is generally used in fitting models of source structure to visibilities. In that case, the $\chi^2$ estimator is typically given as $\chi^2=\bm{\bar{X}}^T\bm{\Sigma_{uc}}^{-1}\bm{\bar{X}}$, where $\bm{\bar{X}}=\bm{\tilde{X}}-\bm{X}$, $\bm{\tilde{X}}$ and $\bm{X}$ represent the \textit{observed} and the \textit{model/true} visibility vectors respectively, and $\bm{\Sigma_{uc}}$ is the uncorrelated data covariance matrix which is only diagonal ($\bm{\Sigma_{uc,ij}}=\delta_{ij}\sigma^2_i$) containing the visibility measurement variances $\sigma^2_i$. This is the general method used inside $\texttt{DIFMAP modelfit}$. However, even if a diagonal visibility covariance matrix is prescribed for the fit, the best-fit parameters may be correlated. A Monte Carlo "bootstrap" method is an effective way to "repeat" the same observation numerous times without writing observing proposals, to understand uncertainties, correlations and bias in estimation of best-fit parameters of the model. It is vaguely similar to FR-RSS (Flux Randomization and Random Subset Selection; \citealt{pet98}) or bootstrapping to understand uncertainties in VLBI rotation measure maps of jets \citep{pasch19}. This can be used further if one uses a pre-defined model in a well sampled $u-v$ plane as a starting point and progressively "worsens" the $u-v$ coverage \textit{randomly}, and then follows the spread or bias in the final parameter estimation using the resulting parameter covariance matrix.

\subsection{Closure quantities}

A type of interferometric observables, called closure quantities, are, to first-order, independent of antenna-based gain errors and hence a robust estimator of the source structure. For example, in the case where $H_{ij}$ is identically equal to unity, an amplitude and phase error in Equation \ref{eq:vis} would translate as:

\begin{equation}
\tilde{V}_{ij}=(1+a_i)(1+a_j)e^{\bm{i}(\phi_i-\phi_j)}A_{ij}e^{\bm{i}\Psi_{ij}}+\epsilon_{ij}(t)
\end{equation}

where $a_i$, $a_j$ are the amplitude errors and $\phi_i$, $\phi_j$ are the phase errors per antenna. The true visibility $V_{ij}=A_{ij}e^{\bm{j}\Psi_{ij}}$ with true amplitude and phase $A_{ij}$ and $\Psi_{ij}$ respectively. If a triangle is formed from three such baselines, the observed "closure" phase around these antennas would be given by the cyclic sum of the \textit{observed} phases:

\begin{equation}
\label{eq:clp}
\tilde{\Psi}_{ijk}(t)=[\tilde{\Psi}_{ij}+\tilde{\Psi}_{jk}+\tilde{\Psi}_{ki}]=[\Psi_{ij}+\Psi_{jk}+\Psi_{ki}]+[\epsilon'_{ijk}]
\end{equation}

which is clearly independent of the phase errors $\phi$ and is a representation of the "true" phase, to a first order in a thermal noise $\epsilon'$ (see \citealt{bb20} for the appropriate statistics). Since $\Psi_{ij}=-\Psi_{ji}$, note that the closure phase is purely antisymmetric in $i,j,k$. In the limit of high SNR, the closure phase variance can be separated out from a diagonal covariance matrix as $\sigma_{clp}$ \citep{bb20}, which is given by \citep{chael18}:

\begin{equation}
\label{eq:var1}
\sigma_{clp}=\sqrt{\sigma_{ij}^2/|\tilde{V}_{ij}|^2+\sigma_{jk}^2/|\tilde{V}_{jk}|^2+\sigma_{ki}^2/|\tilde{V}_{ki}|^2}    
\end{equation}

Among $^{N}C_3$ (which is a binomial coefficient with $n=N=$ number of antennas and $k=3$) closure triangles, a large number is redundant, implying they can be formed from an independent set of closure triangles, which are $^{N-1}C_2$ in number. This implies closure phases encode $(N-2)/N$ fraction of the total information in the complex visibility. As evident, for large $N$ maximum information is retained, which is favourable for $N=27$ for the VLA. However, there are multiple independent sets possible. It is expected that any choice of the independent set should encode the same information in the limit of high SNR and large N \citep{bb20}. A closure phase error, in this regard, is one that cannot be rectified by forming a closure phase. This translates to antenna-independent uncorrelated phase errors $\phi(t)$, which will appear as more additive terms in Equation \ref{eq:clp}.

Similarly, if a closure quadrangle is formed from four baselines, a possible closure amplitude is given by:

\begin{equation}
A_{ijkl}=\frac{|\tilde{V}_{ij}||\tilde{V}_{kl}|}{|\tilde{V}_{ik}||\tilde{V}_{jl}|}=\frac{|V_{ij}||V_{kl}|}{|V_{ik}||V_{jl}|}
\end{equation}

Like closure phases, closure amplitudes have significant redundancy. The total number of such independent amplitudes is given by $N(N-3)/2$ (e.g., \citealt{bb20}). The error per closure amplitude is given by:

\begin{equation}
\label{eq:var2}
\sigma_{\rm cl-amp}=A_{ijkl}\sqrt{\frac{\sigma_{ij}^2}{|\tilde{V}_{ij}|^2}+\frac{\sigma_{kl}^2}{|\tilde{V}_{kl}|^2}+\frac{\sigma_{ik}^2}{|\tilde{V}_{ik}|^2}+\frac{\sigma_{jl}^2}{|\tilde{V}_{jl}|^2}} 
\end{equation}

As evident, multiplicative errors but \textit{specific} to every antenna can be disposed of using this method, making it a very robust estimator of the source structure. The information about the absolute flux and position are of course lost in the process, which can otherwise be obtained from default methods. Therefore, using closure quantities to image or fit models to radio observations is a so-called "calibration-insensitive" way to constrain intrinsic source properties, as used by the EHT team for example \citep[e.g.,][]{chael18}. A possible geometric picture of closure phases can be found in \cite{thyag20}. However, it is also clear that when there are baseline-dependent errors, we shall have additional closure (or "non-closing") errors. For the VLA, it is generally issues in antenna delay settings or non-orthogonality in the sine-cosine correlators \citep{perley99}. Amplitude and phase closure errors are generally expected to be $\lesssim 1\%$ and $\sim$ few degrees, respectively. However, for a real dataset, the scatter in the amplitudes and phases may exceed that expectation for closure errors (due to additional unfixed antenna-based errors) and hence modelling them with a \textit{general} prescription of multiplicative errors as in Equation \ref{eq:ge} may be recommended.

\section{Code Description}
\label{sec:code}
\texttt{DIFMAP}, written in C, is highly modular and has functionalities ranging from plotting the visibility data to deconvolution given in separate source code files. Particularly relevant for this purpose are \texttt{modfit.c} and \texttt{lmfit.c} which are dedicated to model fitting and \texttt{uvf\_read.c} that "reads" in a \texttt{uvfits} dataset. Hence we mainly edited three source codes to suit our purpose in \texttt{ngDIFMAP}. The main edits can be summarized as follows in the corresponding subsections. We edited \texttt{DIFMAP} and did not write a code anew is because of how accepted and well tested it is in the field: a new code necessarily entails a number of disadvantages, including bug fixing and maintaining it over time. In contrast, it was much more suitable to use the engine of $\texttt{DIFMAP}$ and only add/modify functionalities.

\subsection{Closure Quantity Fitting using Simulated Annealing}

Assuming a chosen model using which visibilities are generated and closure quantities are computed, a new function \texttt{getnextclp()} inside \texttt{modfit.c} returns the data minus model residuals for closure amplitudes and phases to the model fitting function inside \texttt{lmfit.c}. We have followed the prescriptions of \cite{chael18} and \cite{bb20} for this purpose. The corresponding likelihood regularizer used is :

\begin{equation}
\begin{split}
\chi^2=\frac{1}{N_{\rm cl-ph}+N_{\rm cl-amp}}\Bigg[\sum_{n=1}^{N_{\rm cl-amp}}\frac{|A_{\rm n,obs}-A_{\rm n,model}|^2}{\sigma_{\rm n,cl-amp}^2}+\\
\sum_{n=1}^{N_{\rm cl-ph}}\frac{|e^{i\tilde{\psi}_{\rm n,obs}}-e^{i\tilde{\psi}_{\rm n,model}}|^2}{\sigma_{\rm n,cl-ph}^2}\Bigg]
\end{split}
\end{equation}

where the two sums are over only independent closure amplitudes and phases respectively with the corresponding subscripts. This of course only works when the closure quantities are negligibly correlated, implying the covariance structure is ignored. Although this can cause severe discrepancies when the number of antennas is low \citep{bb20}, this assumption is safe for arrays like the VLA, where $N=27$. For determining errors, we have employed other techniques as discussed in Section \ref{sec:det_err}. The variances $\sigma_{\rm cl-ph}$ and $\sigma_{\rm cl-amp}$ follow from Equations \ref{eq:var1} and \ref{eq:var2}. However, the corresponding closure amplitude error can diverge for low SNR (Equation \ref{eq:var2}) and deviate significantly from a Gaussian distribution, which will invalidate the used least-squares assumption. This is not as severe for the closure phase errors. Using log closure amplitudes makes the error distribution resemble more of a Gaussian distribution \citep{chael18} and the least-squares estimator can be used. The resulting $\chi^2$ is better written as \citep{chael18}:

\begin{equation}
\begin{split}
\chi_{\rm cl-amp}^2=\frac{1}{N_{\rm cl-amp}}\Bigg[\sum_{n=1}^{N_{\rm cl-amp}}\frac{A_{\rm n,obs}^2}{\sigma_{\rm n,cl-amp}^2}\\
\bigg(\log\bigg|\frac{A_{\rm n,obs}}{A_{\rm n,model}}\bigg|\bigg)^2\Bigg]    
\end{split}
\end{equation}

In the above, it is implicitly assumed that $A_{model}\neq0$, and for $0<A_{model}\ll 1$, the initial choice of model is not very far away from the observation, otherwise a logarithmic divergence is possible. It essentially boils down to practice which one of the above prescriptions may be chosen and the user may choose any inside \texttt{ngDIFMAP}. In addition, inside \texttt{lmfit.c} we added few functions dedicated to "Simulated Annealing" (SA), which is a global optimization algorithm \citep[e.g.,][]{goffe94}. It is a variant of the Metropolis algorithm, where it starts from an initial parameter set and a "temperature" and evaluates the neighbouring parameter space randomly for a number of iterations. For each case, it measures the corresponding Gibbs probability given the observed dataset, given by $p\sim\exp(-\chi^2/T)$. While the solution with the highest probability is chosen, worse solutions are not always rejected (with a threshold) to prevent falling into possible local minima. This algorithm is repeated for a given number of times (or function evaluations) and at each step the temperature drops and the probability of rejections decreases, when the algorithm slowly converges to the correct global minima. The speed and accuracy of this algorithm is dependent on a number of "tuning" parameters, whose description can be found in the code documentation and in \cite{goffe94}. The $\chi^2$ is fed to this optimizer inside $\texttt{lmfit.c}$ and it accordingly searches for the global minima. The $\chi^2$ can be user-described, which can be complex visibilities, or closure quantities. However, since this is a global optimizer, it takes considerably more time than the present Levenberg-Marquadt (LM) optimizer inside \texttt{DIFMAP}. For VLA data, it can take anywhere between one and seven days, which is heavily dependent on the size and simplicity of the dataset. An option for partial parallelization using \texttt{openmp} has been provided in \texttt{lmfit.c} but it only reduces the total time by at most 10\%. Parallelization using the Message Passing Interface (MPI) is not possible without rewriting entire \texttt{DIFMAP}. The user has the option to use any of the above (SA or LM), but the LM optimizer has not been modified to incorporate closure quantity fitting since it requires completely rewriting the existing the fitting module of the code. This will be done in a future endeavour.

\subsection{Editing Visibility Data as desired}

This source code reads the complex visibilities from a \texttt{uvfits} file through function \texttt{get\_uvdata()}. Therefore, the data being fed to \texttt{modelfit} can be directly modified by editing \texttt{uvf\_read.c}. This is the crux of the modification: manipulate the data in any desired way. Furthermore, a few additions inside \texttt{modfit.c} allows to print the model visibilities to an ASCII file, which can then be read through \texttt{uvf\_read.c} in a second run \textit{instead} of the observed uvfits file. We have added this functionality, implying \textit{any} ASCII model visibility can be loaded inside \texttt{DIFMAP} using \texttt{uvf\_read.c}, be modified according to \textit{any} prescription (edit $u-v$ coverage, change normalization, add errors, etc.) and the corresponding effects on the \texttt{clean} image as well as \texttt{modelfitting} can be noted. This opens up a vast sea of possible analyses with \textit{any} kind of interferometric data. This functionality, although very straightforward, is unfortunately not available in any publicly radio data analysis software. For most cases, one needs to develop their own code, like the Event Horizon Telescope (EHT) collaboration, in which case they are not always publicly available \citep[e.g.,][]{themis20}. Our addition to \texttt{DIFMAP} makes the radio data analysis suite complete for any radio astronomer who has visibility data observed by \textit{any} telescope.

\section{Model Datasets and Tests of Bias}
\label{sec:mc}
\begin{figure}
    \centering
    \includegraphics[width=\linewidth]{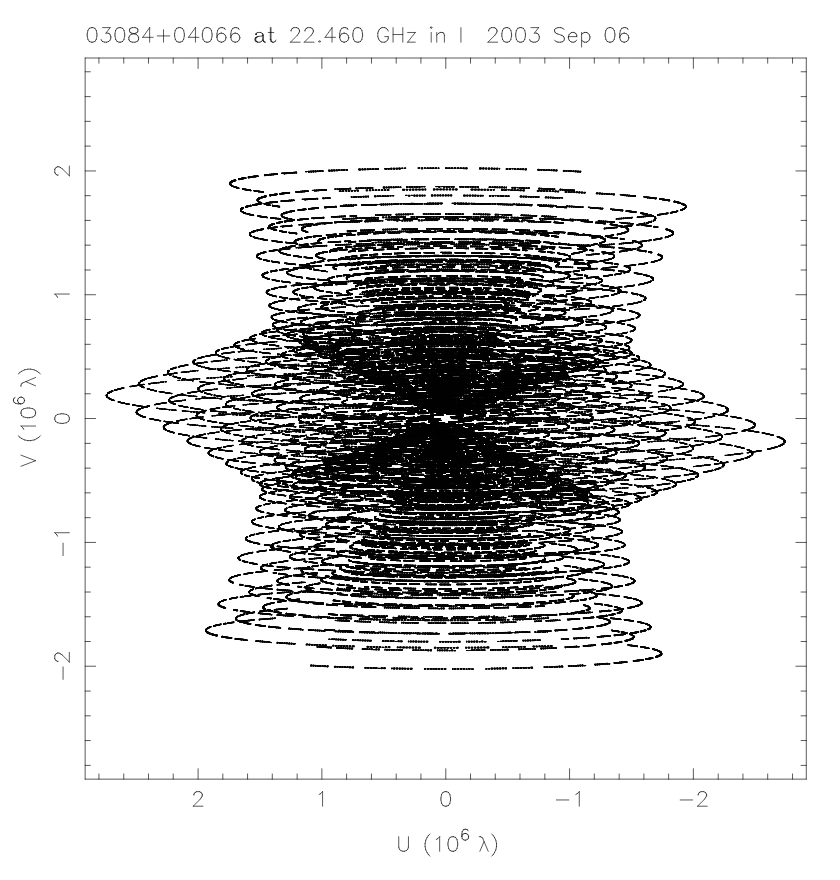}
    \caption{Figure shows the $u-v$ coverage of the 3C 78 22 GHz VLA observation. The corresponding $u-v$ sampling is $\beta_0\simeq3\times10^{-4}/\lambda^2_{2\,cm}$ where $\lambda_{2\,cm}$ is in metres.}
    \label{fig:uv2003}
\end{figure}

The "perfect" model dataset is one with an infinitely well-sampled $u-v$ plane, and an infinite range of spatial frequencies. However, every radio telescope has a "window" of spatial frequencies it can sample and as integration time $T$ on a source or the number of antennas $N$ increases, for the VLA where the minimum thermal noise, or the inverse of sensitivity $s\propto [N(N-1)T]^{-0.5}$, the corresponding rate of increase of sensitivity $1/s$ decreases, as $\partial s/\partial T\propto -T^{-1.5}$ and $\partial s/\partial N\propto -N^{-2}$ for $N\gg1$. Therefore it is only relevant to start from a concrete example of a well-sampled visibility data observed in ideal weather conditions. The tests should be further relevant to the bigger objective of this paper, i.e., to develop a survey of proper motions in VLA observed extragalactic jets. The typical VLA field of view for frequencies less than 50 GHz is given as $\Theta=42/\nu_{GHz}$ arcminutes. This implies for a sample VLA 22 GHz observation, $\Theta\sim2'$, where the expected number of background sources is negligible. The typical length of nearby jets ($z<0.1$) for observations toward the higher end of the frequency range are typically close to few arcseconds. We choose a sample dataset of a 3C 78 22 GHz (2 cm) Very Large Array observation from 2003, which is a dual-channel observation (implying two "Intermediate Frequency" channels at 22 GHz at 22.435 and 22.485 GHz, each having width 50 MHz) and has an angular resolution $\sim0.09"$, with a $\sim2"$ jet at 22 GHz. In order to deal with the effect of changing $u-v$ coverage, we have defined a pertinent prescription $\beta=nD^2/B^2$, where $n$ refers to the number of visibilities at a specified polarization, $D$ and $B$ refer to the antenna diameter and the maximum baseline length respectively. For our given dataset, the number of visibilities per polarization is $n\sim2\times10^{6}$, $D=25$ m and $B\simeq2\times10^{6}\lambda_{2\,cm}$ and therefore the corresponding \textit{default} $u-v$ sampling is $\beta_0\simeq3\times10^{-4}/\lambda^2_{2\,cm}$ where $\lambda_{2\,cm}$ is in metres. Figure \ref{fig:uv2003} shows the $u-v$ coverage of the sample observation, which is visibly very well-sampled. However, even then, "rays", following the antenna pattern, arise in the synthesized beam of this observation when natural weighting is used. Since we are not working in the image plane, we choose the $u-v$ coverage of this observation for our starting model for the rest of the paper. 

In this section, we employ different starting models spanned by the above $u-v$ coverage to create starting synthetic datasets. For each case, we fit the real and imaginary parts of the visibilities and check the behaviour of best-fit model parameters (determined from \texttt{modelfit}) in response to worsening of the $u-v$ coverage and gain errors. Furthermore, the corresponding behaviour is also dependent on the choice of parameters of the starting model, which is also explored.

\subsection{Loss of $u-v$ coverage and the covariance matrix}
\label{sec:uv}

\begin{table*}
\label{tab:par1}
\centering
  \caption{Initial parameter configurations for different Monte Carlo runs in Section \ref{sec:uv}. $d$ and $\theta$ are the corresponding polar coordinates in a left-handed system with the origin at the phase tracking centre. The Major FWHM and axial ratio refer to that for the Gaussian. The position angle (P.A.) measures an effective tilt to the East from the vertical at the phase centre. The relative flux values, and not absolute flux densities, are relevant since this is not a real observation. Therefore the point source parameters have been fixed and the parameters of the knot/Gaussian are tuned through various ranges specified below. No ranges are provided for values that have been kept fixed.}
  \begin{tabular}{cccccccc} \hline\hline 

Component & Flux (F) & Distance (d) & $\theta$ & Major FWHM (w) & Axial Ratio & P.A. \\
& & (mas) & (degrees) & (mas) & (mas) & (degrees) \\
\hline
Point Source & 1 & 0 & 0 & --- & --- & --- \\
Gaussian & $0.01, 0.035, 0.1, 0.5$ & $200, 400, 600$ & 45 & $50, 100, 200$ & $0.8$ & 0 \\
\hline
\end{tabular} 
\end{table*}

We start from a simple model of a jet that consists of a bright point source in addition to a knot modelled by a two-dimensional Gaussian. We use a range of parameters for the core and the knot, tabulated in Table \ref{tab:par1}, and for each corresponding configuration, we study possible correlations between the parameters and their variances by creating $N=1000$ realizations per given $u-v$ coverage. To achieve this, we start from the initial $u-v$ plane sampling and using a uniform random number generator inside C (\textit{rand()}), generate a fully new subset for each realization (modifying the seed every time using the computer's internal clock) and then follow the best-fit parameters from the fitting. While more complex and robust algorithms to generate random numbers exist, we believe the technique described above must provide a good approximation for this work. We describe this entire setup using a parameter covariance matrix, where we check the dependence of the off-diagonal as well as the diagonal (variance) terms on the $u-v$ coverage and choice of the initial starting model for different sets of parameters. We found $u-v$ sampling $\beta\gtrsim0.1\beta_0$ to have negligible effect on the best-fit parameters and therefore we only discuss effects of worse sampling. We therefore initialize runs with $u-v$ coverage corresponding to $0.1\beta_0$ (90\% removal) and $5\times10^{-4}\beta_0$ (99.95\% removal), which we found to cover the entire gamut of $u-v$ coverage for our specific model placed with sampling $\beta_0$ in a VLA 22 GHz observation. For the rest of the paper we shall refer to the $u-v$ coverage as $\beta$.

\begin{figure*}
    \centering
    \includegraphics[width=0.9\linewidth]{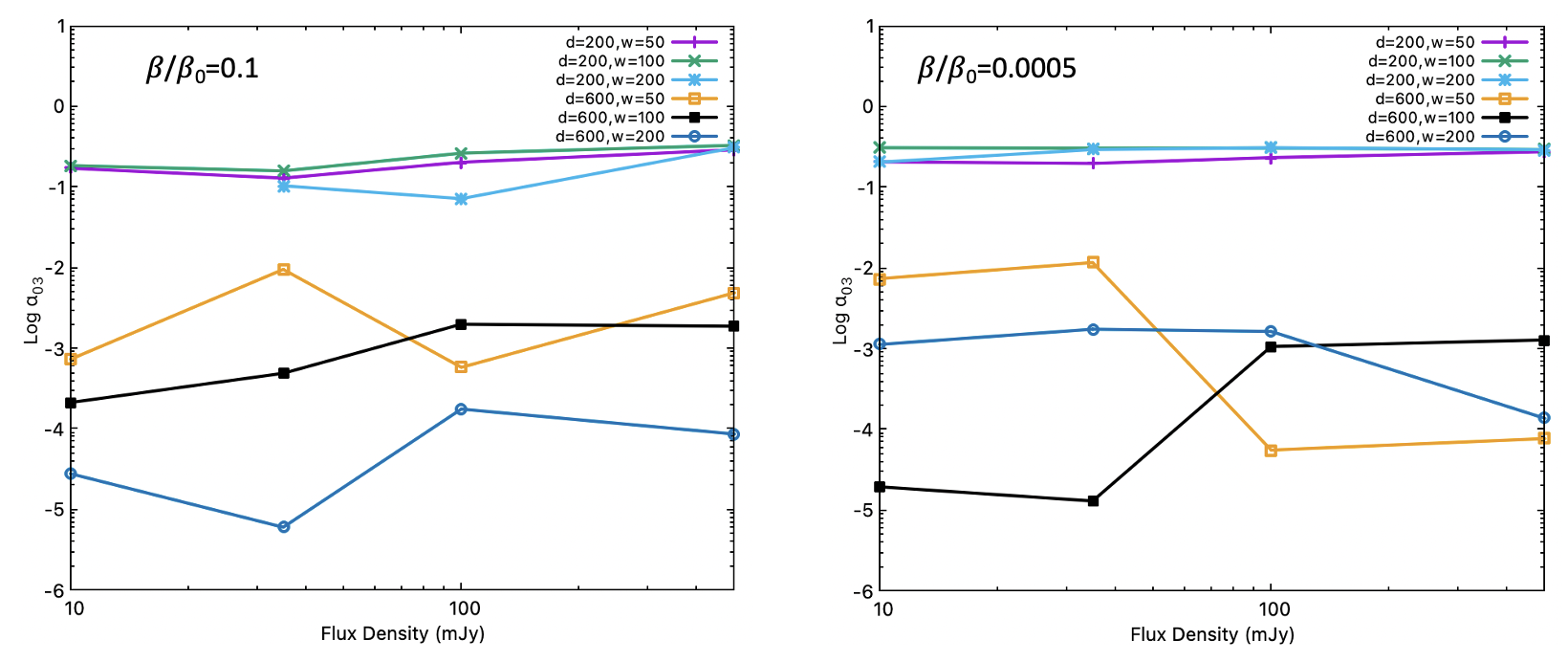}
    \caption{Figure shows $\alpha_{03}$, or the square of the correlation between the core flux (0) and the Gaussian flux density (3), in log-scale v/s Gaussian flux density for different parameter configurations, namely for $d=200, 600$ and major axis $w=50, 100, 200$ and for different $u-v$ coverage $\beta$. Left: $\alpha_{03}$ v/s Gaussian flux density for $u-v$ coverage $\beta/\beta_0=0.1$. Right: $\alpha_{03}$ v/s Gaussian flux density for $u-v$ coverage $\beta/\beta_0=0.0005$. Both the panels show strong correlation $\alpha\sim0.1$ when the distance to the core $d$ is smaller ($d=200$), which is marginally higher for worse $u-v$ coverage, but also mostly independent of the Gaussian flux. The correlation can be intuitively understood as a consequence of progressive "resemblance" of the Gaussian to the core, which affects the fitting algorithm.}
    \label{fig:corr03_uvonly}
\end{figure*}

A measure of the strength of the off-diagonal terms between two parameters with indices $(m,n)$ in the resulting \textit{parameter covariance matrix} $\Phi$ can be given as $\alpha_{mn}=(\Phi^2_{mn})/(\Phi_{mm}\Phi_{nn})\in(0,1)=r_{mn}^2$ where $r_{mn}$ is the Pearson correlation coefficient. Greater correlation between the two parameters will result in the higher values of $\alpha_{mn}$. However, it is by default unclear what value of $\alpha$ would be considered statistically significant. For this, we use the $t-test$ statistic, which is given by $t=r\sqrt{N-2}/\sqrt{1-r^2}$ \citep[e.g.,][]{bowley_std_rho,rahman1968}, where $N$ is the number of degrees of freedom, which is $1000$ in our case. The use of this t-test statistic is justified because of a large sample size ($N\sim1000$) and the random variables follow a normal distribution (see Figure \ref{fig:all_ge}). The minimum $r_{mn}$ to produce a p-value $<0.05$ turns out to be $|r_{mn}|\gtrsim r_{min}\simeq0.065$ (for a two-tailed test), implying $\alpha_{mn}\gtrsim4\times10^{-3}$ for statistically significant correlation. We define the parameter indices as follows: Point Source Flux Density (0), $X_{PS}$ (1), $Y_{PS}$ (2), Gaussian Flux Density (3), $X_G$ (4), $Y_G$ (5), Major Axis (6), where $X$ and $Y$ refer to the corresponding locations of the component in the Eastern and Northern directions from the phase centre respectively. We plot $\alpha_{mn}$ for the pairs (0,3), (3,4), (3,6), (4,5) and (4,6) in Figure \ref{fig:corr03_uvonly} in the main paper, and Figures \ref{fig:corr34_uvonly}, \ref{fig:corr36_uvonly}, \ref{fig:corr45_uvonly} and \ref{fig:corr46_uvonly} respectively in Appendix \ref{sec:app_vbc}, for $\beta/\beta_0=0.1$ and $\beta/\beta_0=0.0005$.

Figure \ref{fig:corr03_uvonly} shows $\alpha_{03}$, or the square of the correlation between the core flux (0) and the Gaussian flux density (3) v/s Gaussian flux density for different parameter configurations and for different $u-v$ coverage $\beta$.  Interestingly, both the panels show strong correlation $\alpha\sim0.1$ when the distance to the core $d$ is smaller ($d=200$), which is marginally higher for worse $u-v$ coverage, but also mostly independent of the Gaussian flux. The correlation is negative and it can be intuitively understood as a consequence of progressive "resemblance" of the Gaussian to the core, which affects the fitting algorithm. The other cases have negligible correlation for the same reason.

From the marginal probability distributions for selected parameters (figure omitted for brevity), the median value of the best-fit parameter distribution is chosen, which equals the mean in all of our simulations. The mean/median best-fit parameter is chosen in this way and we find there is $\lesssim10^{-6}$ fractional bias in estimating the same for both $\beta$, implying a \textit{perfectly calibrated} VLA observation of a simple extragalactic jet at poor $u-v$ coverage $\beta$ but $\beta\gtrsim10^{-4}\beta_0=3\times10^{-8}/\lambda^2_{2\,cm}$ will still produce the same mean parameter values if it had $\beta=\beta_0=3\times10^{-4}/\lambda^2_{2\,cm}$.

\begin{figure}
    \centering
    \includegraphics[width=0.9\linewidth]{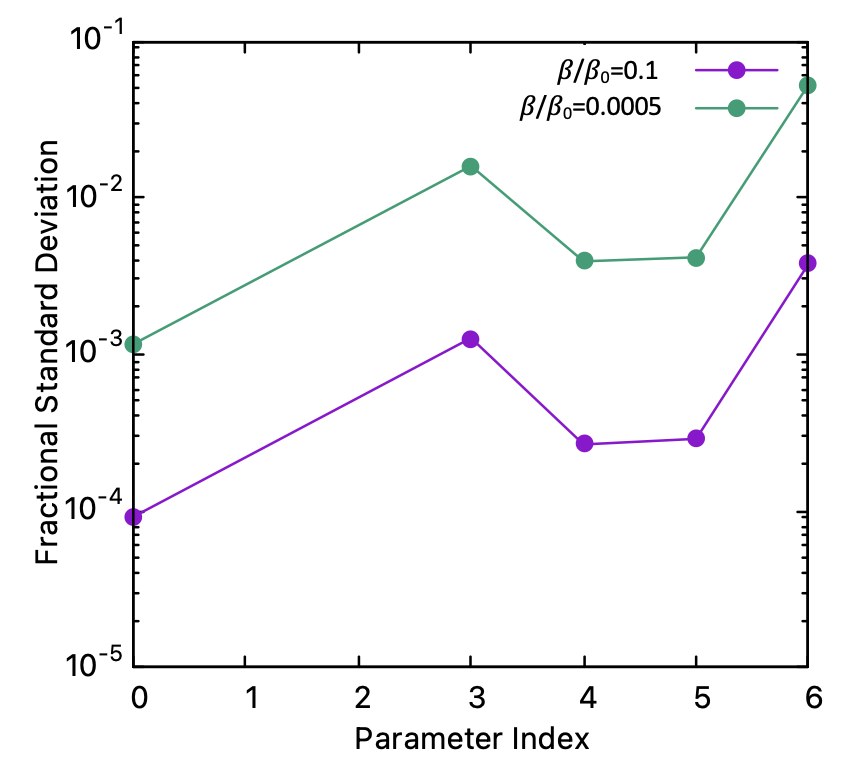}
    \caption{Figure showing the fractional parameter standard deviation averaged from the diagonals of all the covariance matrices of each parameter configuration, as a function of parameter index for different $u-v$ coverage $\beta$. The uncertainties are typically less than 10\% and show values higher by an $\sim$ order of magnitude for $\beta/\beta_0=0.0005$}.
    \label{fig:sig}
\end{figure}

The figures discussed above and in Appendix \ref{sec:app_vbc} show the extent of correlation between various model parameters in the simple case of a least squares fit assuming uncorrelated visibilities. However, in \textit{most} cases even though the correlation is statistically significant, it is particularly poor and much less than the maximum observed correlation coefficient $r_{max}\simeq0.3$, which is only mild. This implies that in most simple jet structures the diagonal elements of the covariance matrix shall suffice as lower limits to the variances in the event of absence of correlated closure errors. The fractional uncertainties from the covariance matrix diagonal have been plotted for all the parameters for different $\beta$ in Figure \ref{fig:sig}. For $\beta/\beta_0=0.0005$, the uncertainties are higher as expected and they are found to be $\lesssim$ an order of magnitude. In spite of the off-diagonal nature of the parameter covariance matrix, the variance estimate is particularly simple for any function that is \textit{derived} from the parameters. An extremely relevant case is that of the X and Y positions of the Gaussian, which are otherwise only used as relative to the core. This implies the measured quantity is $\Delta X=X_1-X_0$ (or same for Y), where $X_0$ and $X_1$ are the East positions of the core and the Gaussian with respect to the phase centre. Therefore $\sigma^2_{\Delta X}=\sigma^2_{X_0}+\sigma^2_{X_1}-2\sigma_{X_0X_1}$. This works as an approximate lower limit to the standard deviation on $\Delta X$ (or $\Delta Y$) since this is only due to imperfect $u-v$ coverage without presence of gain errors.

In this subsection, we have not discussed systematic removal of baselines due to specificity in the external cause of antenna dropouts. There can be infinite possibilities of proceeding with such a method and is only relevant when the effect on model fitting due to removal of specific antennas needs to be determined. This can, however, be carried out easily inside \texttt{uvf\_read.c} according to any prescription the user needs.

\subsection{Simulating Gain/Closure Errors}
\label{sec:ge}





\subsubsection{Preliminaries}

If for a set of independent true visibilities $V_{ij}\sim \mathcal{N}(x_k,\sigma_k)$ ($0\leq k\leq n$, $n$ represents total number of visibilities), the gain factors in $G_i$ or $H_{ij}$ in Equation \ref{eq:vis} are real and $\sim\mathcal{N}(1,\sigma_g)$ at all times are independent and identically distributed, $\{\tilde{x}_1$,...,$\tilde{x}_i$,...,$\tilde{x}_N\}$ will remain independent. However, a term dependent on $x_i$ will appear as modification to the diagonal of the default covariance matrix of the visibilities. It is straightforward to show that the modified $\Sigma_{uc,ij}=(\sigma^2_i+\sigma^2_i\sigma_g^2+\sigma_g^2x_i^2)\delta_{ij}$. In presence of additional phase errors distributed normally $\sim\mathcal{N}(0,\sigma_p)$, the matrix is similarly modified with $\Sigma_{uc,ij}=(\sigma^2_i+\sigma^2_i\sigma_t^2+\sigma_t^2x_i^2)\delta_{ij}$, where $\sigma_t^2=\sigma_g^2+\sigma_p^2$.

\begin{figure}
    \centering
    \includegraphics[width=\linewidth]{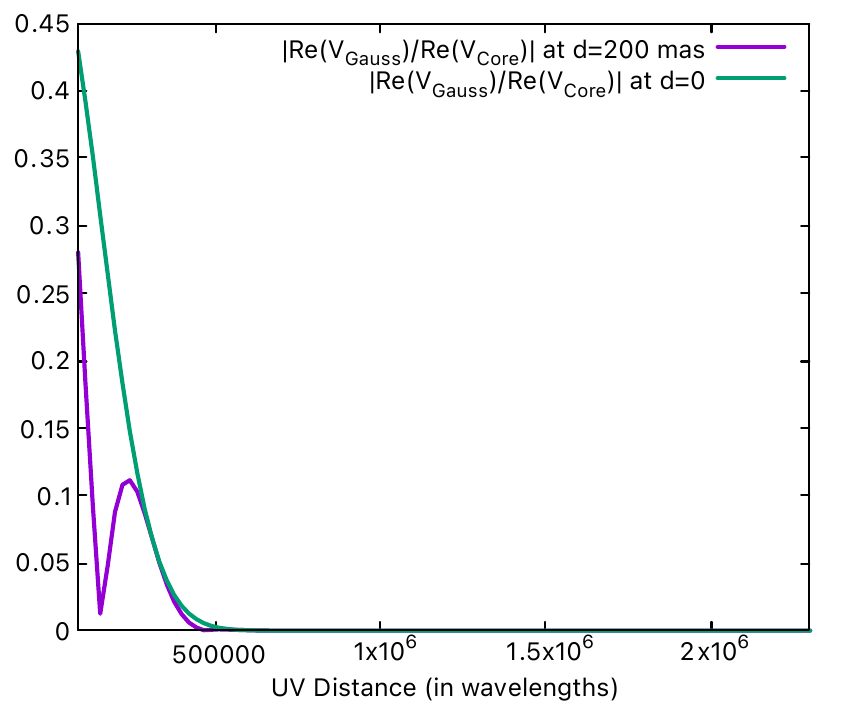}
    \caption{$|Re(V_{\rm gauss})/Re(V_{\rm core})|$ v/s UV distance for $d=200$ mas and $d=0$ plotted between baseline lengths $B_{\rm min}=8.6\times10^{4}\lambda_{\rm 2\,cm}$ and $B_{\rm max}=1.8\times10^6\lambda_{\rm 2\,cm}$ at 22 GHz VLA A configuration. The cosine term for $d>0$ strongly suppresses the visibility amplitude.}
    \label{fig:vgauss}
\end{figure}

As long as the closure errors are independent and identically distributed across all visibilities, the covariance matrix will be diagonal. However, the diagonal elements will have dependence on the visibilities themselves and hence it must be taken into account during least-squares model fitting to provide unbiased estimates of the fit parameters. For cases when the source is core-dominated but has extended emission, as in the case of an AGN with a jet (modelled by a single Gaussian), it can be shown that $Re(V(u,v)_{ij})=[F_{\rm core}+F_{\rm gauss}f_{ij}\cos(ux+vy)]$ (where $F$ denotes the flux density and $f_{ij}=e^{-\Gamma^2/ln\,2}$) and $Im(V_{ij})=F_{\rm gauss}f_{ij}\sin(ux+vy)$ (core at phase center). $\Gamma=a\sqrt{u^2+v^2}$ for a two-dimensional Gaussian of width $a$, where both the P.A. and axial ratios are 0 and 1 respectively. Since  $e^{-\Gamma^2/ln\,2}<1$ and decreases \textit{rapidly} with $u-v$ distance, even when $F_{\rm gauss}=0.5F_{\rm core}$ (like several cases in our model in Table \ref{tab:par1}), $[Re(V_{ij})]^2\sim F^2_{\rm core}$ and $[Im(V_{ij})]^2\sim 0$, with a stronger effect for larger widths. To verify this, we have plotted in Figure \ref{fig:vgauss} $|Re(V_{\rm gauss})/Re(V_{\rm core})|$ for $d=0, 200$ mas, $w=50$ mas and $F_{\rm gauss}=0.5F_{\rm core}$, where the cosine term for $d=200$ strongly suppresses the visibility amplitude as opposed to $d=0$. It can be shown that the total length through which $|Re(V_{\rm gauss})/Re(V_{\rm core})|\geq10\%$ is $\sim54057\lambda_{\rm 2\,cm}$, which is \textit{only} $\sim 3\%$ of the total UV distance covered between baseline lengths $B_{\rm min}=8.6\times10^{4}\lambda_{\rm 2\,cm}$ and $B_{\rm max}=1.8\times10^6\lambda_{\rm 2\,cm}$. This implies that in the case where our model Gaussian flux density is maximum, the diagonal elements of the real part of the covariance matrix are effectively modified by an added constant $\sigma^2_g F^2_{\rm core}$, while the imaginary part remains unchanged. Therefore further modification of the covariance matrix was not required for any of the used models where the Gaussian flux density even lower. However, if required, the covariance matrix can be accordingly modified to include measurement-dependent diagonal elements inside ngDIFMAP.

\subsubsection{Effects of simulated errors on correlations and variance}

The previous discussion evidently assumes a priori knowledge of the gains. As discussed in the next subsection, an estimate of the gain error can be derived by attempts to mimick the original visibility data. When the knowledge of the possible gain error is obtained or known beforehand, the modification to the least squares estimator is absolutely necessary, except in cases of highly core-dominated structures. The provision for modification is provided inside \texttt{uvf\_read.c}, \texttt{lmfit.c} and \texttt{modfit.c}. 

\begin{figure*}
    \centering
    \includegraphics[width=\linewidth]{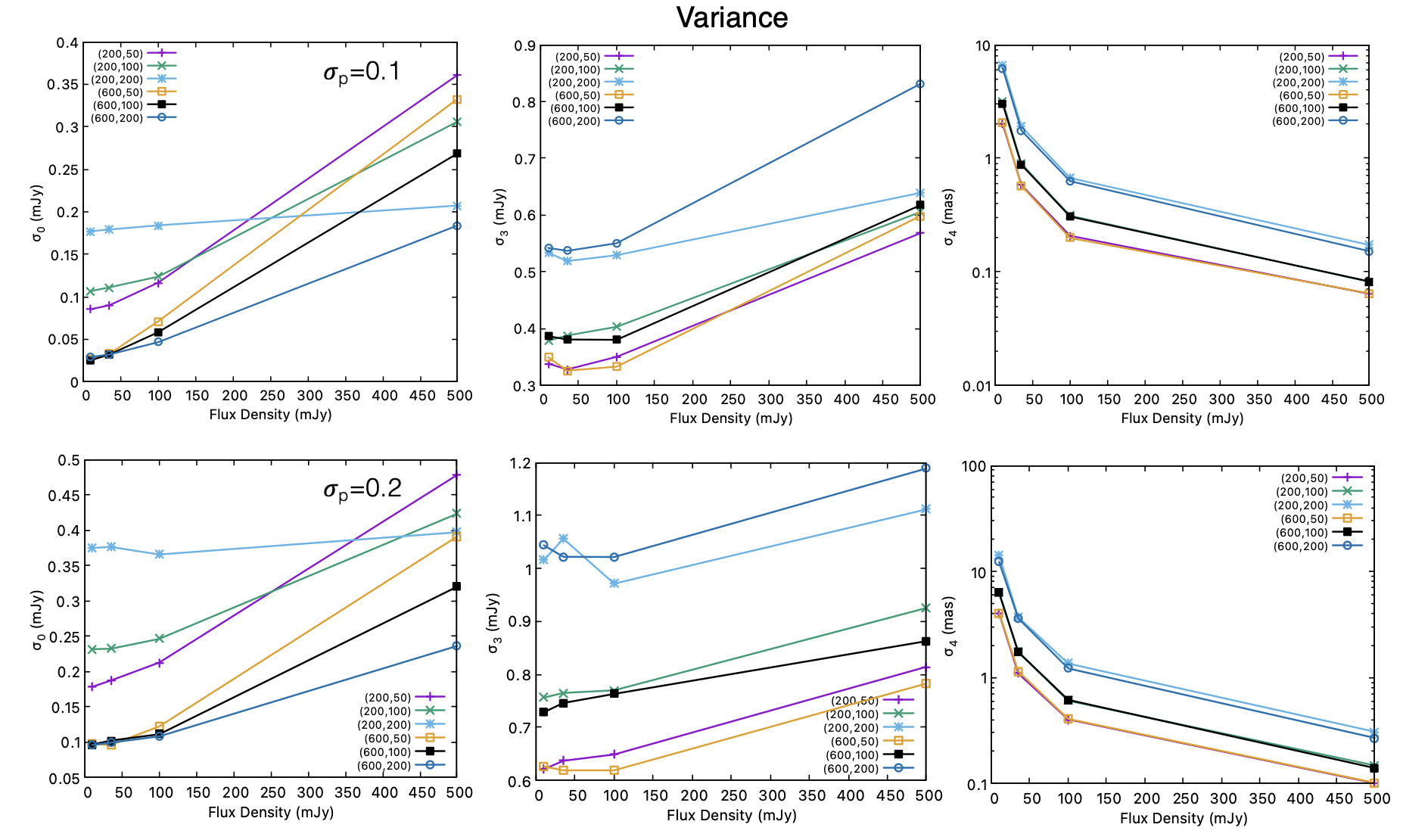}
    \caption{Figure shows the standard deviations of the determined best-fit point source flux (0), Gaussian flux density (3) and the Gaussian X position (4), for different degrees of phase error $\sigma_p$, as a function of the Gaussian flux. The legend shows the various configurations in a format (distance from the core, width of the Gaussian), or $(d,w)$. Top panel: For 0.1 radian phase error. Both $\sigma_0$ and $\sigma_3$ increase with increase in Gaussian flux density, implying a constant fractional variance. However, their strengths are similar for both phase errors and on an average lower than that for amplitude errors. Similarly $\sigma_4$ decreases as expected with increase in Gaussian flux density and is the lowest for the most compact component. However, $\sigma_4$ is much larger than than that in amplitude errors, which is typical of a phase error since it affects component position more than flux or size. Bottom panel: For 0.2 radian phase error. The behaviour is very broadly similar to the top panel, except the uncertainties are 1-2 times larger. The faintest Gaussian closest to the point source and largest in size shows higher uncertainties on average, as it comes "under the influence" of the point source. $\sigma_4$ is larger than the top panel since phase error is higher.}
    \label{fig:pe_sig}
\end{figure*}

In this section, we simulate the effect of amplitude and phase errors on the best-fit parameters of our initial model given in Table \ref{tab:par1} for a given $u-v$ coverage $\beta/\beta_0=0.1$. This is done using a diagonal measurement covariance matrix with almost no dependence on the real or imaginary visibility data, or with a constant diagonal. This approximation holds as long as a bright point source dominates most of the source emission. We simulate 1000 realizations following previously described methods, and for each prescription of gain error, we plot the corresponding correlations among parameters 0, 3, 4 and 6, as well as the bias and variances of parameters 0, 3 and 4. We keep only a subset of our results for the main paper section, while all the others and their further details can be found in Appendix \ref{sec:app_vbc}. Figures \ref{fig:ge_corr}, \ref{fig:pe_corr}, \ref{fig:gepe1_corr}, \ref{fig:gepe2_corr} contain the correlations, while Figures \ref{fig:pe_sig}, \ref{fig:ge_sig}, \ref{fig:gepe1_sig}, \ref{fig:gepe2_sig} contain the variances and Figures  \ref{fig:pe_bias}, \ref{fig:ge_bias}, \ref{fig:gepe1_bias}, \ref{fig:gepe2_bias} contain the biases produced in parameter estimation for different prescriptions of gain error.

Figures \ref{fig:ge_corr}, \ref{fig:pe_corr}, \ref{fig:gepe1_corr} and \ref{fig:gepe2_corr} show the square of the correlations between the point source flux (0) and the Gaussian flux density (3) ($\alpha_{03}$), Gaussian flux density (3) and Gaussian X position (4) ($\alpha_{34}$), Gaussian flux density (3) and Gaussian width (6) ($\alpha_{36}$) for different degrees of amplitude and phase error ($\sigma_g$ and $\sigma_p$ as defined in Equation \ref{eq:ge}), as a function of the Gaussian flux. We choose this specific set of parameters since they are more commonly inferred for physical conclusions. The legend shows the various configurations in a format (distance from the core, width of the Gaussian), or $(d,w)$.

In Figure \ref{fig:ge_corr}, for 5\% amplitude error, all the correlations, $\alpha_{03}$, $\alpha_{34}$ and $\alpha_{36}$ are statistically significant. The largest correlation through all the three cases is dominated by the configuration with the Gaussian closer to the core and having a larger width, in which case the fitting routine finds it increasingly difficult to separate the core from the Gaussian for non-zero amplitude error. In the case of the 15\% amplitude error, the situation is similar for the Gaussian closer to the core, while that farther away (d=600) has developed more correlations on an average, and particularly so for $\alpha_{36}$.

\begin{figure*}
    \centering
    \includegraphics[width=\linewidth]{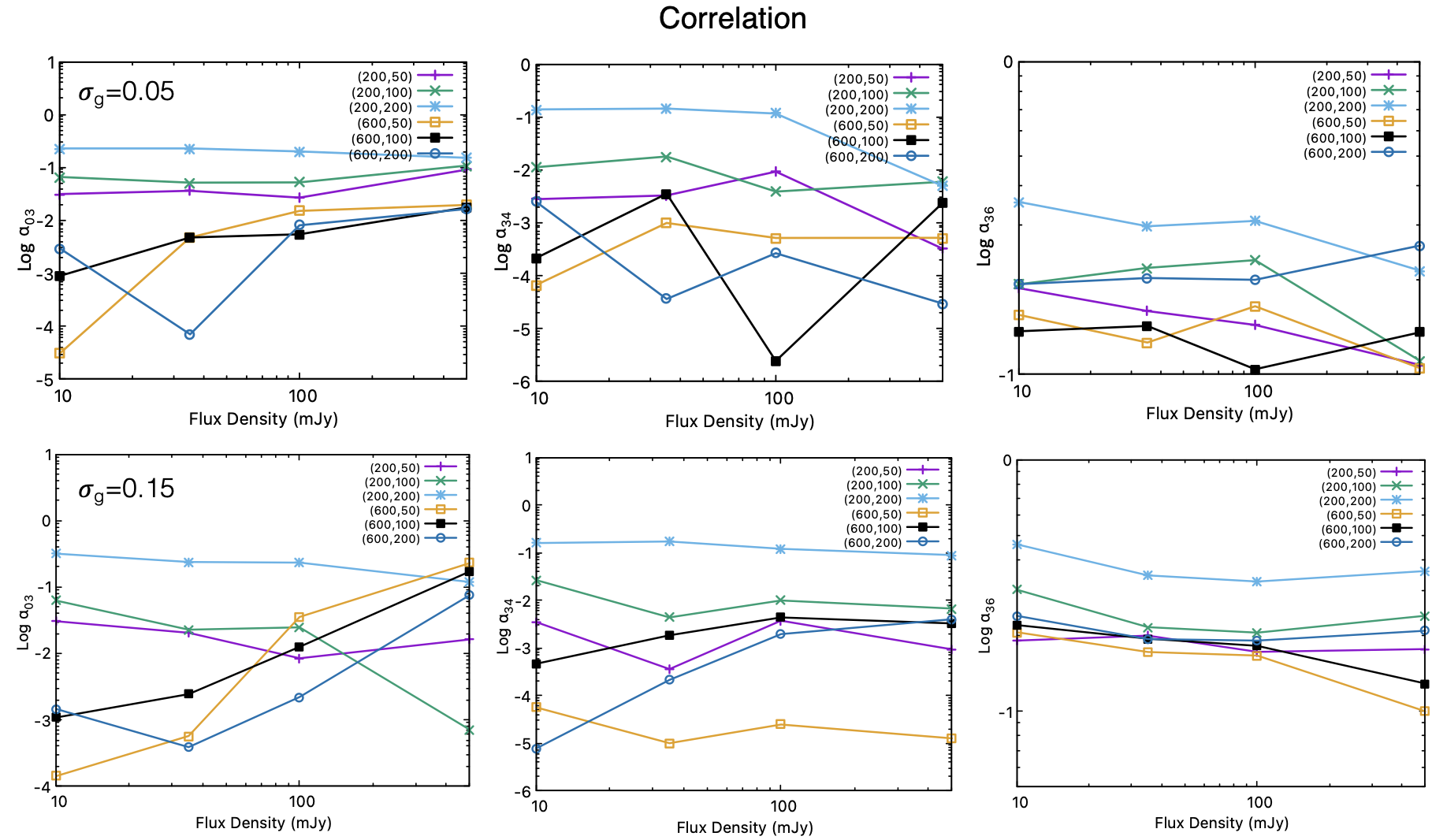}
    \caption{Figure shows the square of the correlations between the point source flux (0) and the Gaussian flux density (3) ($\alpha_{03}$), Gaussian flux density (3) and Gaussian X position (4) ($\alpha_{34}$), Gaussian flux density (3) and Gaussian width (6) ($\alpha_{36}$) for different degrees of amplitude error $\sigma_g$, as a function of the Gaussian flux. The legend shows the various configurations in a format (distance from the core, width of the Gaussian), or (d,w). Top panel: For 5\% amplitude error. All the correlations, $\alpha_{03}$, $\alpha_{34}$ and $\alpha_{36}$ are $\sim$ 1 dex larger than that in the absence of random amplitude error. The correlations have become statistically significant, and the largest through all the three cases is dominated by the Gaussian configuration closer to the core and a larger width.
    Bottom panel: For 15\% amplitude error. For all the cases, the correlations for the $(200,w)$ case have mostly remained similar to the 5\% case, while they have strengthened for the Gaussian at d=600 from the core.}
    \label{fig:ge_corr}
\end{figure*}

Figures \ref{fig:pe_sig}, \ref{fig:ge_sig}, \ref{fig:gepe1_sig}, \ref{fig:gepe2_sig} show the standard deviations (or $\sqrt{\Phi_{mm}}$) for the point source flux (0), Gaussian flux density (3) and Gaussian X position (4), for the same prescriptions of amplitude and phase error. The legend refers to the corresponding $(d,w)$ pair, as mentioned before.

In Figure \ref{fig:pe_sig}, for both 0.1 radian (6 degrees) and 0.2 radian phase error, $\sigma_0$ and $\sigma_3$ show systematic linear increase with Gaussian flux density, implying a constant fractional variance, like previous figures. $\sigma_0$ is higher for Gaussian components closer to the core and larger. Additionally, the more compact Gaussians have lower $\sigma_3$ on an average for the same Gaussian flux. However, both $\sigma_0$ and $\sigma_3$ are very similar for both low and high phase errors and also lower than that for amplitude errors, which is partly expected since phase errors generally affect the component positions only. This is confirmed by larger values of $\sigma_4$ compared to the amplitude error case, which is highest for the 0.2 radian phase error. Like Figure \ref{fig:ge_sig}, the brighter and more compact component can be better localized, or its $\sigma_4$ is lower. It should be noted that for either of amplitude or phase errors, the variances are considerably larger than that in absence of them, when compared to Figure \ref{fig:sig}.

\subsubsection{Effects of simulated errors on bias}

Figures \ref{fig:pe_bias}, \ref{fig:ge_bias}, \ref{fig:gepe1_bias}, \ref{fig:gepe2_bias} show the bias in estimating the point source flux (0), Gaussian flux density (3) and Gaussian X position (4), for the same prescriptions of amplitude and phase error. The legend refers to the corresponding $(d,w)$ pair. The bias for $x$ is simply given as $(x_{\rm obs}-x_{\rm expected})/x_{\rm expected}$.

In Figure \ref{fig:pe_bias}, for 0.1 and 0.2 radian phase errors we see roughly similar behaviour for biases for parameter indices 0, 3 and 4. Both the point source flux and Gaussian flux densities are heavily negatively biased, by almost -1\% and -3\% for the lower and higher phase errors respectively. The Gaussian X position ($Bias_4$) is also negatively biased and more so for the higher phase error (order of magnitude increase) and the Gaussian component \textit{farther away} from the core. While the negative bias in X position can be understood to be the effect of the core dominance, the reason why components with larger $d$ are more negatively biased and $Bias_3$ and $Bias_4$ are negative is more subtle. This can be explained using a very simple toy model of visibility in one coordinate $u$, with a core (flux $F_{\rm core}$) and a Gaussian (flux $F_{\rm gauss}$) of $u$ coordinate width $\sim1/\sqrt{a}$ located at $x_0$ in the image plane. The model visibility can be written as:

\begin{figure*}
    \centering
    \includegraphics[width=\linewidth]{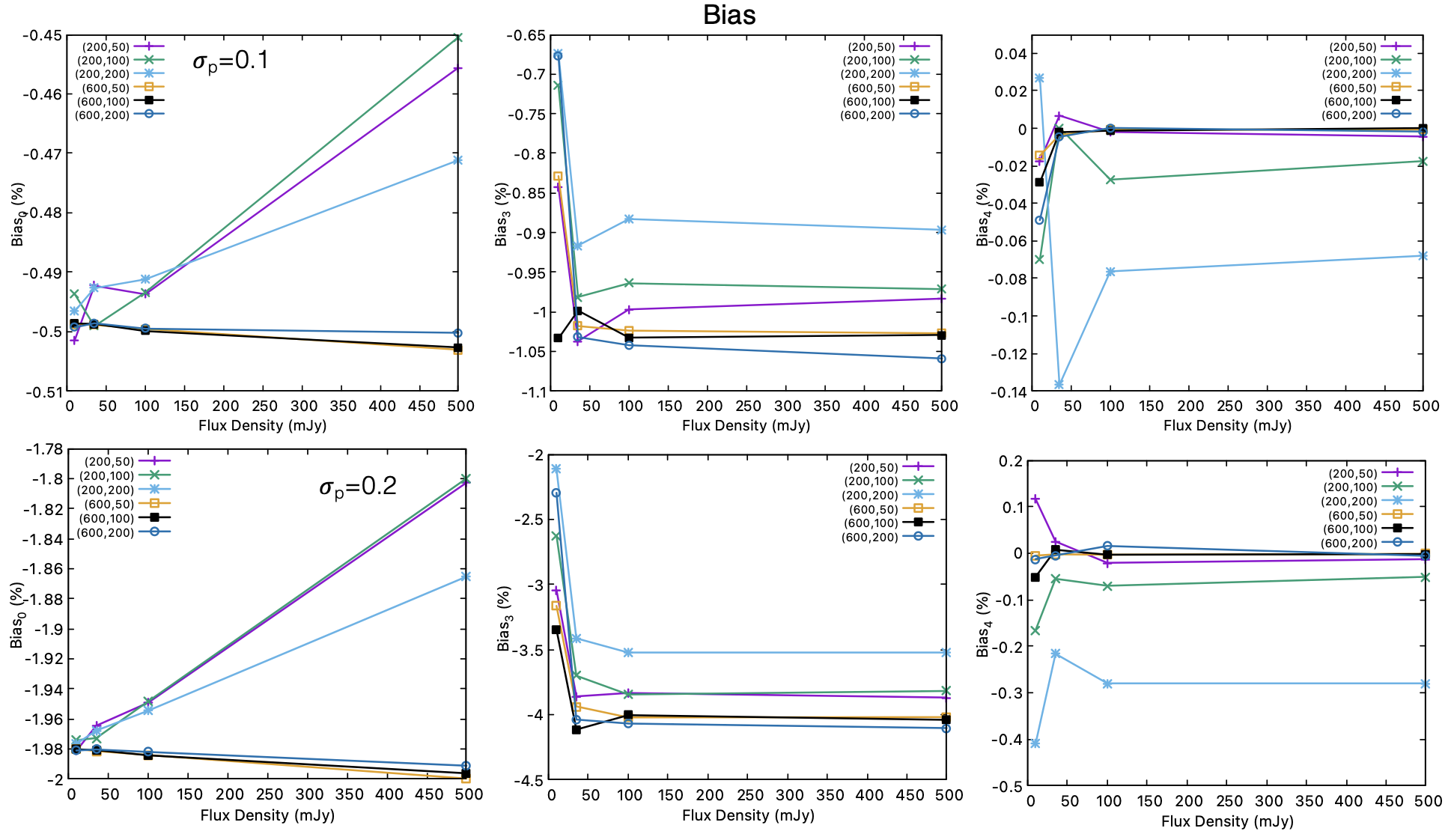}
    \caption{Figure shows the biases of the determined best-fit point source flux (0), Gaussian flux density (3) and the Gaussian X position (4), for different degrees of phase error $\sigma_p$, as a function of the Gaussian flux density. The legend shows the various configurations in a format (distance from the core, width of the Gaussian), or $(d,w)$. Top panel: For 0.1 radian phase error. $Bias_0$ and $Bias_3$ are strongly negative and more so for components far away from the core. $Bias_4$ is also negative. All of this is because addition of a phase error systematically lowers the visibility amplitude and more so for fast oscillating components, or those with high $d$. Bottom panel: For 0.2 phase error. While all the biases are very similar in shape to the top panel, they have increased by 5-10 times. }
    \label{fig:pe_bias}
\end{figure*}

\begin{equation}
V_{\rm mod}(u)=F_{\rm gauss}e^{-au^2+\bm{i}ux_0}+F_{\rm core}
\end{equation}

The "observed" visibility will be modified by phase errors $\sim\delta\phi\in [-\pi/2,\pi/2]$:

\begin{equation}
\begin{split}
V_{\rm obs}(u)&=(F_{\rm gauss}e^{-au^2+\bm{i}u x_0}+F_{\rm core})e^{\bm{i}\delta\phi}\\
&=(F_{\rm gauss}e^{-au^2}\cos(ux_0+\delta\phi)+F_{\rm core} \cos\delta\phi)+\\ &\bm{i}(F_{\rm gauss}e^{-au^2}\sin(ux_0+\delta\phi)+F_{\rm core} \sin\delta\phi)
\end{split}
\end{equation}

This directly implies $Re(V_{\rm obs})<Re(V_{\rm mod})$ and $Im(V_{\rm obs})<Im(V_{\rm mod})$ on an average. For $F_{\rm core}=0$, $\Delta V=V_{\rm obs}-V_{\rm mod}=-2F_{\rm gauss}e^{-au^2}\sin(ux_0+\delta\phi/2)\sin(\delta\phi/2)+\bm{i}(2F_{\rm gauss}e^{-au^2}\cos(ux_0+\delta\phi/2)\sin(\delta\phi/2))$. To have an idea of the overall behaviour of $\Delta V$ with respect to $x_0$, the quantities to look at would be $A=Re(\partial \Delta V/\partial x_0)=-2F_{\rm gauss}e^{-au^2}x_0\cos(ux_0+\delta\phi/2)\sin(\delta\phi/2)$ and $B=Im(\partial \Delta V/\partial x_0)=-2F_{\rm gauss}e^{-au^2}x_0\sin(ux_0+\delta\phi/2)\sin(\delta\phi/2))$. We note that the presence of the exponential damping term ($e^{-au^2}$ from the Gaussian) heavily suppresses $Re(\Delta V)$ and $Im(\Delta V)$ at $u$ beyond the first sign flip for both A and B, which implies the both the quantities retain the initial same sign on averaging over $u$. Therefore, both A and B would be negative quantities with all later sign flips having an extremely reduced effect due to the damping. Hence $A, B\propto -x_0$ and this causes only underestimation of the Gaussian and core flux densities for components far from the core and hence we see more negative bias with increase in $x_0$ in Figure \ref{fig:pe_bias}. Additionally, in the limit of $F_{\rm core}\gg F_{\rm gauss}$, $V_{\rm obs}\simeq F_{\rm core}\cos\delta\phi+\bm{i}F_{\rm core}\sin\delta\phi$. If $\delta\phi\sim\mathcal{N}(0,\sigma_p)$, the effect of $\sin\delta\phi$ averages out due to equal positive and negative contributions. In contrast, $\cos\delta\phi$ is symmetric and systematically positive in $[-\pi/2,\pi/2]$, implying it stays as a factor less than unity. $V_{\rm obs}$ is therefore smaller for larger $\sigma_p$, implying a model fit will systematically lower the best-fit flux values for both the Gaussian and the core, creating a negative bias increasing with $\sigma_p$ as we see here. Furthermore, while $Bias_3$ tends to a constant value with increase in Gaussian flux density, $Bias_4$, for most components, on an average goes to zero in the limit of large flux.

\subsubsection{Section summary}

\begin{figure*}
    \centering
    \includegraphics[width=\linewidth]{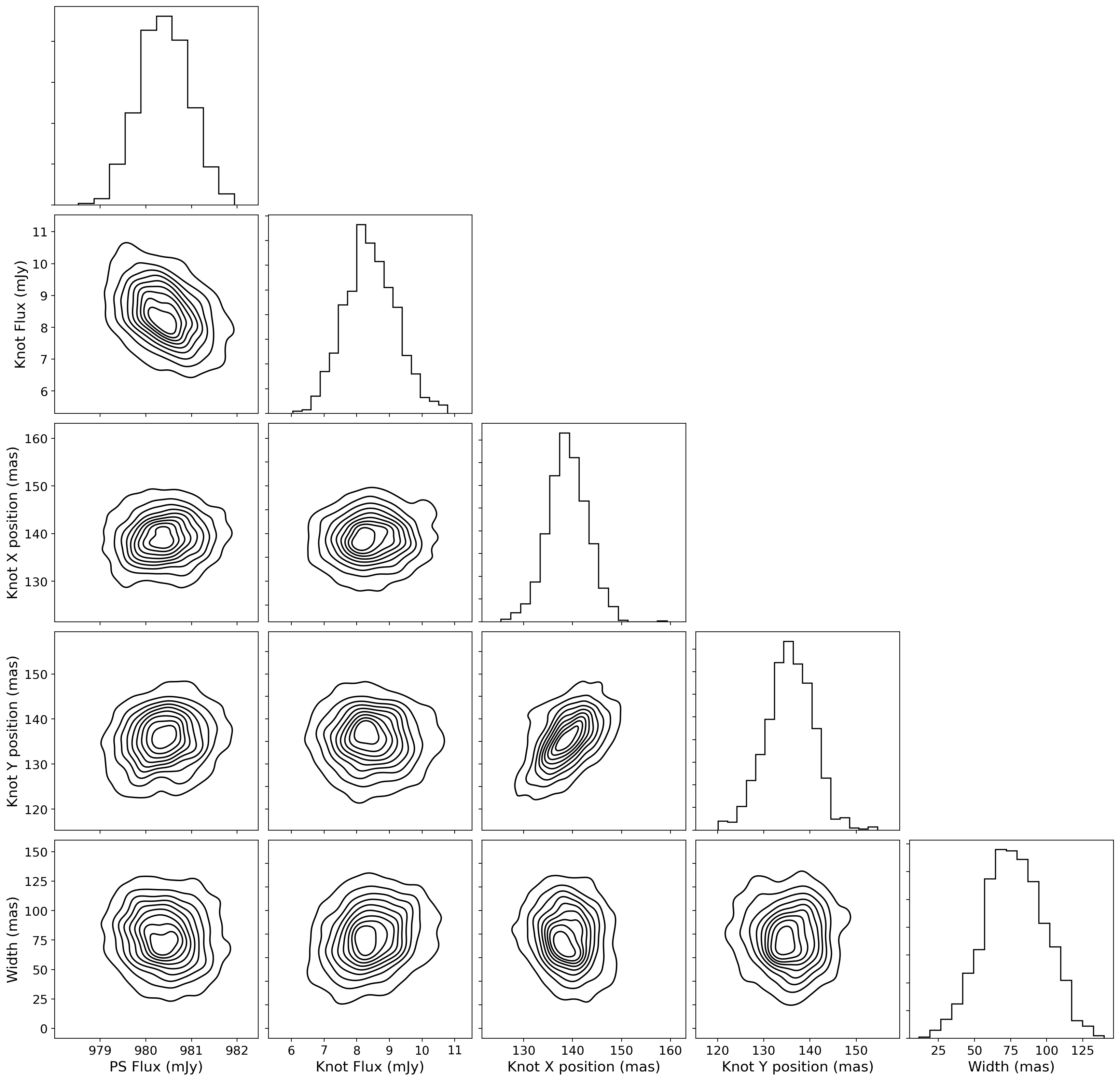}
    \caption{Marginal probability distributions for selected parameters obtained from bootstrapping, for configuration with distance of the Gaussian to the core d=200, width of the Gaussian w=200, flux density of the Gaussian F=10 mJy, $u-v$ coverage $\beta/\beta_0=0.1$, amplitude error $\sigma_g=15\%$ and phase error $\sigma_p=0.2$ radian. Since the scales are not similar, it is difficult to discern correlations if there is any. The Gaussian X and Y positions are positively correlated, while the Gaussian and point source flux densities are negatively correlated. Contours were made from a kernel density estimator and are \textit{not} representative of any confidence region.}
    \label{fig:all_ge}
\end{figure*}

Figure \ref{fig:all_ge} is one representation of the marginal probability distributions of the parameters and possible correlations. From first look, the point source and Gaussian flux densities are clearly negatively correlated, while we see a positive correlation between X and Y Gaussian positions. Since $\tilde{V_{ij}}\sim (1+Y_{ij})e^{-a(u^2+v^2)}\cos(uX+vY)$, the multiplicative amplitude errors should affect the determination of parameters inside the cosine. $X$ and $Y$ must be affected \textit{equivalently}, thereby creating a correlation, with the degree and dominant sign of correlation dependent on $u$ and $v$ (and the $u-v$ coverage \textit{pattern}), the coefficients of X and Y. In other cases, a correlation, even if present, may not be evident since the X-axis and Y-axis scales are not always similar. Therefore, looking at the $\alpha$ figures is the most clear way of detecting possible correlations. We also note that although bootstrapping does not generally guarantee that the marginal distribution must be Gaussian, they are strongly symmetric, and mostly resembling a Gaussian distribution. This implies one can interchangeably use the mean and median of the distribution in this case to quote the best-fit mean value and directly use the variance. The situation is similar in the absence of gain errors, which has not been shown here for more clarity. 


\begin{figure*}
    \centering
    \includegraphics[width=\linewidth]{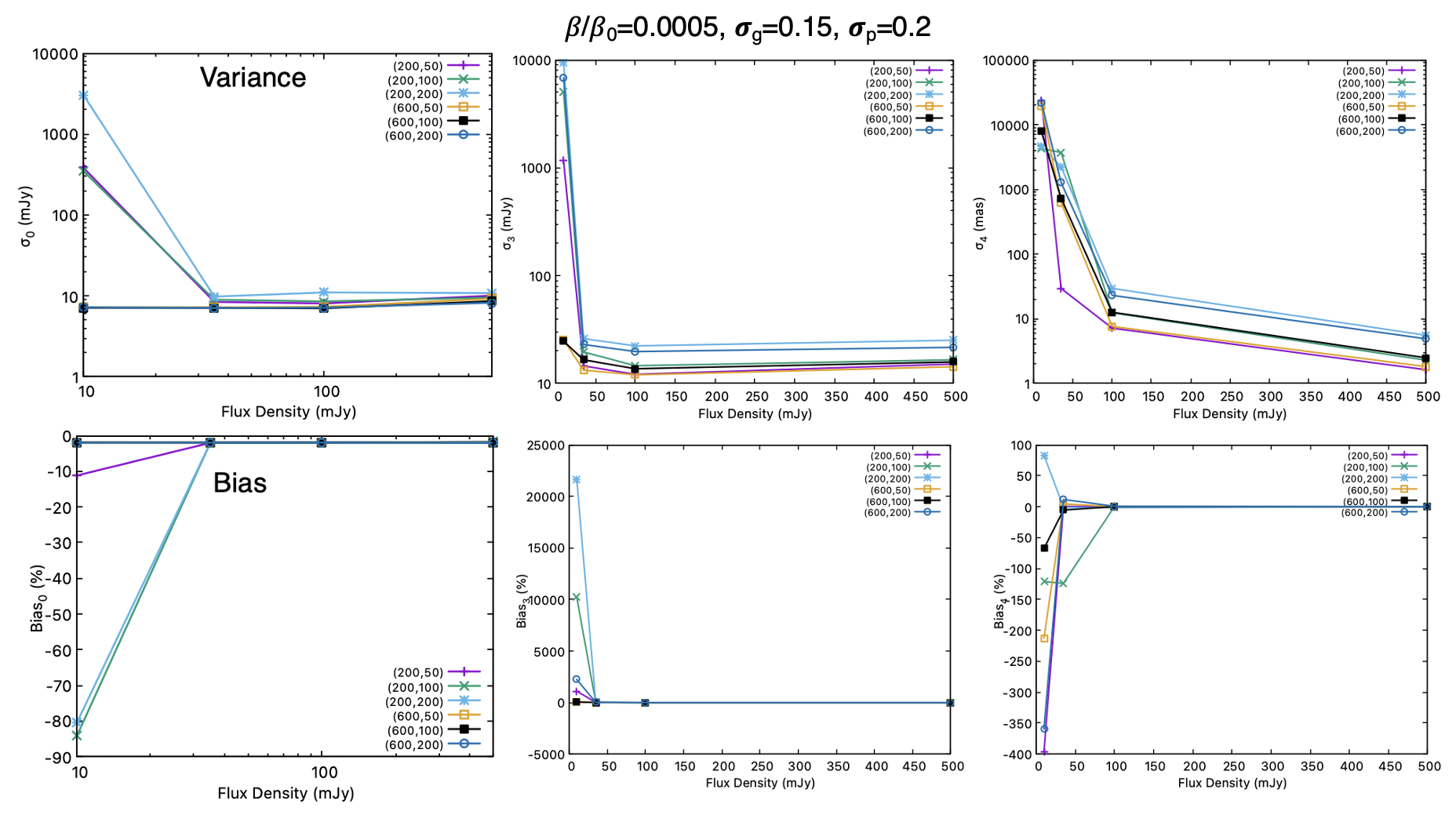}
    \caption{Variances and biases of parameters 0, 3 and 4 at $u-v$ coverage $\beta/\beta_0=0.0005$, amplitude error $\sigma_g=15\%$ and phase error $\sigma_p=0.2$ radian. All of them are absurdly high $\sim$ orders of magnitude at lower Gaussian flux densities. This shows that the data is practically unusable if the Gaussian is not bright enough.}
    \label{fig:gepe2_baduv}
\end{figure*}

All the above analyses were done with fair $u-v$ coverage, or $\beta/\beta_0=0.1$. However, for $\beta/\beta_0=0.0005$, the biases and variances become absurdly large for the worse case of the amplitude and phase error. Figure \ref{fig:gepe2_baduv} shows the same. While the flux and positional standard deviations are thousands in mJy and mas respectively, the bias is between few tens and few hundred percent for the lower flux cases. It is clear that while such a poor $u-v$ coverage by default works, in a real high resolution VLA observation of an extragalactic jet that has combined amplitude and phase errors, the dataset can be essentially considered irretrievable. The limiting $\beta$ until which substantial information can still be recovered is dependent on the source structure and the $u-v$ coverage \textit{pattern}. This will be thoroughly explored for VLA and especially VLBI in a future work. In the limit of highest Gaussian flux density, the biases and variances are much lower, as expected. This result shows that the biases and variances are more heavily affected by $u-v$ coverage worsening when there are already amplitude and phase errors present.

\subsection{Estimating Errors in Real Datasets}
\label{sec:det_err}

\begin{figure*}
    \centering
    \includegraphics[width=0.8\linewidth]{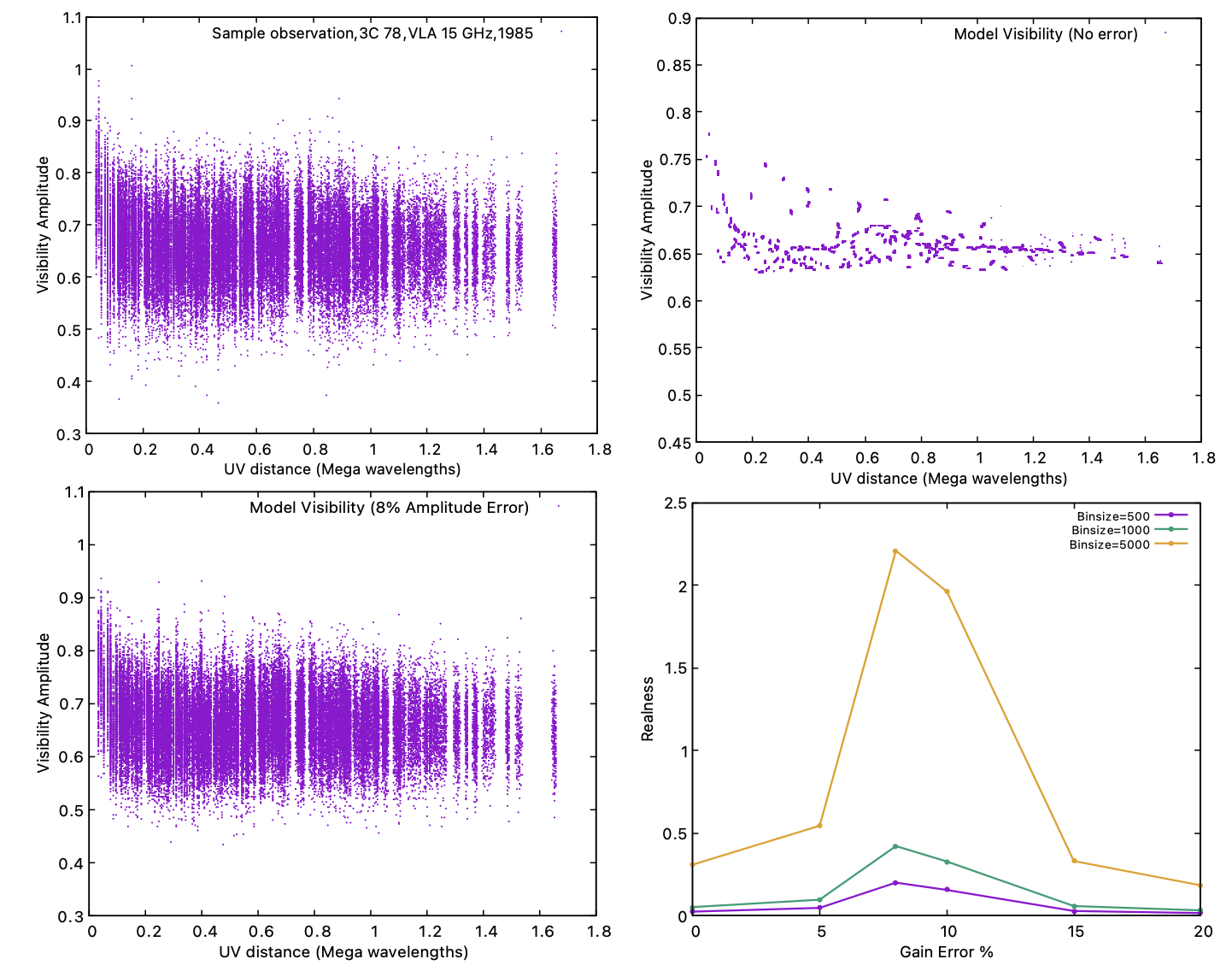}
    \caption{Figure shows how a synthetic dataset can be made noisy to represent a real dataset. Top left: Visibility amplitude v/s UV distance for a VLA 15 GHz observation of radio galaxy 3C 78. Top right: Corresponding best-fit model of the data. Bottom Right: Values of the Realness parameter between the model visibility amplitude and the observed visibility amplitude for different strengths of amplitude error. It peaks at 8\% amplitude error. Bottom left: The synthetic dataset (model) \textit{after} application of 8\% amplitude error, which looks very similar to the real visibility amplitudes, on an average.}
    \label{fig:realness}
\end{figure*}

The previous subsections only discussed the effects of errors and $u-v$ coverage on model fitting visibilities; however, they need to be applied to real datasets. The core idea to mimick a real dataset is to model its possible errors. While it is very difficult from first principles since the antenna complex gains are a priori unknown, Equation \ref{eq:ge} rather provides a simple prescription to provide a rough estimate of the same. Assuming that the real dataset is best described by a specific model, the deviations from the model, or essentially the average scatter of the dataset, must encode the average property of the errors it is being afflicted by. While a degree of correlation/coherence between visibilities may be expected due to phase/amplitude errors, that would require provision of two extra parameters, which would be the magnitude of the correlation and the correlation timescale. Introduction of this realism would add even more dimensions to our analysis that already has a very large parameter space, and we hence chose Equation \ref{eq:ge} as the \textit{simplest} possible prescription to mimick the scatter in real data. If the real and model+error datasets look \textit{similar on average}, our results should be modified negligibly. \cite{natarajan17} use a Bayesian model-selection approach to fitting the antenna gains in addition to the model parameters, in order to determine the extent of VLBI super-resolution in presence of errors. The provision to do this inside \texttt{ngDIFMAP} is available, where one needs to modify \texttt{modfit.c} accordingly by adding an extra gain parameter to existing models. For this work, we estimate the amount of gain error by applying gain errors to a starting model and gauging the similarity in scatter between the final worsened model and the data. In this regard, we define a "realness" parameter $R$, of any synthetic dataset, that characterizes how close its \textit{average} visibility scatter is to a given observed dataset for the same science target. It can be defined as $R=\sigma_{\rm real}/(\sigma_{\rm real}-\sigma_{\rm model})$, where $\sigma$ is the standard deviation per bin in the total visibility data. This quantification is strictly visual and is also based on the fact that real data are more "noisy", and thus more scattered. Figure \ref{fig:realness} demonstrates this idea. It shows a sample visibility amplitude data of 3C 78 observed at 15 GHz in 1985 with the VLA, which shows considerable scatter. The best-fit model for the source structure is also given on the right, which when acted upon by amplitude errors of $\sim8\%$, begins to resemble the observed visibilities, on an average. This is verified by the plot of the Realness parameter, which peaks at $\sigma_g\sim8\%$.

The similar method can be used for the visibility phases and the corresponding $\sigma_p$ determined. Equipped with an estimate of the possible $\sigma_g$ and $\sigma_p$, it is then straightforward to begin with the synthetic dataset for the starting model, apply corresponding errors and inspect the variance, bias and correlation plots. The bias can be then used to predict how deviant the best fit model parameters are from the intrinsic source description, while the variance will provide an estimate of the spread in the best-fit parameter. In contrast, since the correlations between different parameters are also dependent on the errors, they also must be quoted when mentioning the true best-fit parameters. The only real caveat in this approach is the assumption of a model for the antenna/baseline gain errors. While it will most likely not affect the variances, it is a possible exercise to determine the dependence of the biases and correlations on the prescription of the gain error. 


\section{Discussion}
\label{sec:disc}
In the previous sections we have discussed the effects of gain/closure errors on default parameter estimation from model fitting interferometric data. The core ideas extend to any form of complex variable, where amplitude and phase errors may arise. Although some of the results were expected from analytical estimates, the specific degree to which the parameters may be affected was essentially unknown and untested prior to this attempt for the used type of VLA datasets. However, the cases above were only for a simple jet model of a core and a Gaussian. The results may very quickly become analytically intractable in case of many more components, with different flux densities, sizes and positions. In that case, this is the only way to determine biases, variances and correlations in the parameters. Therefore, the fact that this suite of Monte Carlo simulations can be carried out in simple steps makes \texttt{ngDIFMAP} important. 

We also specifically note that the correlations between parameters discussed in the previous section are only artificial and an artifact of the fitting procedure, independent of any physical scenario. The correlations may reduce or increase depending on the fitting procedure; for example while fitting closure quantities, the correlations in parameters created by antenna-specific gain errors may be reduced. Furthermore, if either of visibility amplitudes or phases are being fit, the biases due to phase or amplitude errors will reduce respectively.


For the larger goal of our project CAgNVAS, we are mainly interested in the positional evolution of components with time. Measurement of proper motions essentially implies tracking the position of the given jet component through epochs. This is akin to following the prescription in Section \ref{sec:det_err} for every epoch, and creating a set of the best-fit positions of the core and the Gaussian with the mention of possible correlations, \textit{after} incorporating for the biases. Generally, on kpc-scales, since one need not worry about VLBI core-shifts, one measures the \textit{relative} position of the Gaussian with respect to the core for each epoch, and as mentioned before if $\Delta X=X_1-X_0$ (where $X_0$ and $X_1$ are the Gaussian and core X positions), $\sigma^2_{\Delta X}=\sigma^2_{X_0}+\sigma^2_{X_1}-2\sigma_{X_0X_1}$. The same steps can be taken for $\Delta Y$. This gives a clear picture on how to accurately measure the positional change of a component over epoch, ignoring $N>2$-order correlations. 
If one is instead interested in quoting only a best-fit parameter (random variable $X$) from model fitting interferometric data, one must quote the following: the mean value, the spread or the variance and the correlations.

Furthermore, a possible caveat in this respect is that all the above analyses have only assumed that the source structure of the \textit{real} dataset must be \textit{exactly} described by the best-fit model, implying the obtained variances, biases and correlations are \textit{only} due to antenna/baseline gain errors. Although this provided a clear picture of the same without being affected by the bias in choice of a model, it is not very advisable to use this for source structures that are visibly non-Gaussian, and cannot be modelled by the suite of options inside \texttt{DIFMAP} or \texttt{ngDIFMAP}. For example, for nearby jets with complex morphology, the dissonance between simple models and real structure is evident as it is difficult to match the observed structure using analytical models. In pc-scale jets observed using the VLBI, simple Gaussian models are often suitable. In contrast, instabilities in large-scale matter-dominated jets due to jet-environmental interactions create shocks of complex geometry, in many cases oblique (see \citealt{bick96,hardee82} for the example of M87). In jets where the speeds are high enough to be worked on in the image plane, complex morphology can be dealt with sophisticated techniques, as in \cite{biretta95}. In contrast, if the jet is very slow, and the structure is visibly very complex, it is indeed very difficult to robustly check for proper motions. In that case, one may need to design more complex models taking inspiration from hydrodynamic simulations of jets. This is a possible future exercise, in which we may add more complex jet models inside \texttt{ngDIFMAP}. We also note that while the situation regarding source structure is rather hopeful for non-nearby jets, the speed accuracies are much worse.

\section{Conclusions}
\label{sec:conc}
This paper is the first article from the CAgNVAS project, which aims to create a catalogue of proper motions of AGN jets using the Very Large Array (VLA) archive. Motivated by the need for accurately measuring VLA proper motions and associated biases, we have presented $\texttt{ngDIFMAP}$, created by adding closure quantity fitting, a global optimizer and added functionalities to $\texttt{DIFMAP}$ to make parameter estimation from model-fitting in the $u-v$ plane more robust. Particularly, we have demonstrated in this paper the effects of interferometric errors on the visibility data and subsequent parameter estimation, which were not demonstrated in any previous publication. The major difference between our new code and possibly similar software written by the EHT is that the latter is only designed for lower volumes of data, spread over a much smaller angular area of sky, albeit at a higher resolution. Instead, $\texttt{DIFMAP}$ is by default written in C and runs very efficiently for almost any interferometric data. We have capitalized on this feature to add new functionalities, which would be useful to any radio astronomer.

Our results can be summarized as follows:

1. \texttt{ngDIFMAP} has been introduced, which can model fit interferometric data using closure quantities, using a global optimizer. The optimizer can also be used to fit visibility data. Furthermore, the code can be easily used to simulate any \textit{user-desired} effects on either the observed visibilities, closure quantities, \textit{or} any synthetic dataset. We have utilized this feature heavily in this paper to determine the ramifications of antenna/baseline gain errors on the best-fit parameters obtained using model-fitting.

2. We have used a point source + Gaussian model to test the effects of removal of $u-v$ coverage and amplitude and phase errors on the thereafter estimated best-fit parameters. For each case of a $u-v$ coverage+error prescription, we used the Monte Carlo technique to understand the statistics from 1000 realizations. A $u-v$ coverage prescription is described by $\beta=nD^2/B^2$ where $n$ is the number of visibilities, $D$ is the antenna diameter and $B$ is the longest baseline length. $\beta_0$ is the default starting $u-v$ coverage, and the effect of $\beta$, as progressively lower fractions of $\beta_0$, on the best-fit model is tested.

3. In case of no errors, we find statistically significant correlations between most of the parameters. However, they are particularly low and only substantial for a few cases. Among them, the point source flux density and Gaussian flux densities are negatively correlated for the Gaussian closer to the point source and larger in size. A negative correlation is also found between the Gaussian flux density and size, which is due to the fitting algorithm trying to "conserve" the visibility amplitude. When $u-v$ coverage is particularly worsened by random removal of points ($\beta/\beta_0=0.0005$), the correlations more or less remain the same, but the Gaussian flux density and size correlations appear stronger on an average.

3. Removal of $u-v$ points also worsens the corresponding parameter variance by roughly an order of magnitude in $\beta/\beta_0=0.0005$, compared to $\beta/\beta_0=0.1$.

4. For both cases of $u-v$ coverage, we observe $<10^{-6}$ fractional bias in estimation of the parameters, implying for most datasets without any error, removal of $u-v$ points until a certain limit will not create bias in the best-fit parameters. 

5. Addition of gain errors made the analyses more realistic, since real datasets are always plagued by the same. For the following points, it is implied we have used amplitude errors $\sim5-15\%$ and phase errors $\sim6-12$ degrees.

6. We find that amplitude errors particularly increase correlations between the Gaussian flux density, size and the point source flux densities compared to no amplitude errors. This is more pronounced for the cases when the Gaussian is large and closer to the point source. 

7. We find that phase errors increase correlations, similar to amplitude errors, but the correlations between the Gaussian flux density and its position are particularly very weak. These results are expected since amplitude errors affect the flux and size of a component more than the position, while it is vice versa for phase errors. 

8. We applied combined amplitude and phase errors and we find the behaviour is dependent on the dominant form of the error. For correlations between point source flux density, Gaussian flux density and Gaussian size, they are much higher than the discrete cases, since both \textit{add} to the correlations. It is low between the Gaussian flux density and position, mainly because phase errors caused decorrelation. These correlations are, on an average, large when the Gaussian component is closer to the core and larger in size.

9. The corresponding variances are similar for both amplitude and phase errors. In separate as well as combined cases, we find that variances of the point source and Gaussian flux densities increase $\sim$ linearly with increase in Gaussian flux density, implying a constant fractional variance. The variances are generally higher when the amplitude and phase errors are larger and when the Gaussian is much faint, large and close to the core. Also as expected, phase errors specifically caused higher variance of the Gaussian position, particularly when it is faint and large. The variance increases with increase in both amplitude and phase error. Additionally, for increasing Gaussian flux densities, the positional variance of the Gaussian decreases. The positional variances are mostly $\gtrsim10$ mas for the faintest component, which is pretty large if one wishes to track its positional evolution with time.

10. We obtain intriguing results for the parameter bias in presence of amplitude and phase errors. The biases increase in magnitude for all the chosen parameters with increase in amplitude and phase error: while it is generally $1\%$, it may reach $\sim$ few to 10 \% when the Gaussian is faint. The biases in the point source flux density are consistent with zero in absence of phase errors. Biases in the Gaussian flux densities are generally on the positive side when phase errors are absent, when the Gaussian is closer to the core. However, addition of phase errors causes increased negative bias in the flux densities, which can not reach zero even in the limit of large Gaussian flux. This has been explained using a very simple analytical consideration. The situation for the Gaussian position is considerably unclear. We find that it may be both negatively and positively biased at lower values of Gaussian flux densities when closer to the core and larger in size. In the limit of high flux density, it tends to zero for \textit{all} the cases, unlike the biases in the point source and Gaussian flux densities. 

11. For much worse $u-v$ coverage ($\beta/\beta_0=0.0005$) at 15\% amplitude and 12 degree phase errors, the data practically becomes increasingly unusable when the Gaussian is faint, closer to the core and larger in size. The variances and biases are $>50\%$ at least, but they decrease considerably in the limit of high Gaussian flux. We will explore this especially for VLBI observations in a detailed upcoming work.

12. Overall, for the specific case of measuring proper motions, one needs to suitably understand the variance and bias in estimating the component position. In cases when the component is faint and close to the core, care needs to be taken; otherwise a false shift of $10$ mas over 2 epochs spread over 10 years at a redshift $z=0.03$ can imply an apparent speed $\beta_{\rm app}\sim1.5c$, when the component may not have even moved at all.

13. A real dataset can be very approximately mimicked if the prescription of Equation \ref{eq:ge} is followed, where that amplitude or phase error is chosen when the synthetic dataset's visibility scatter is similar to that for the real dataset. This shall allow to effectively determine the corresponding variances and biases by using Monte Carlo simulations, like this work.

\texttt{ngDIFMAP} hasn't been made public yet since it requires further structuring for a public distribution. We will provide the code and the documentation on a personal request basis until then.

\section*{Acknowledgements}

ARC and ETM thank the anonymous referee whose comments helped improve manuscript greatly. ARC thanks Zsolt Paragi and Markos Georganopoulos for insightful comments that helped improve the paper, and Akram Touil for enlightening discussions on variable correlations. ARC and ETM acknowledge National Science Foundation (NSF) Grant 12971 that supported this work.

\section*{Data Availability}

The inclusion of a Data Availability Statement is a requirement for articles published in MNRAS. Data Availability Statements provide a standardised format for readers to understand the availability of data underlying the research results described in the article. The statement may refer to original data generated in the course of the study or to third-party data analysed in the article. The statement should describe and provide means of access, where possible, by linking to the data or providing the required accession numbers for the relevant databases or DOIs.



\bibliographystyle{mnras}




\appendix

\section{Variance, Bias and Correlations}

This appendix describes the majority of results of variances, biases and correlations between various parameters of the starting core+jet model as described in Table \ref{tab:par1}, and must be read as continuation of the discussions in Sections \ref{sec:uv} and \ref{sec:ge} in the main text. The parameter indices are as follows: Point Source Flux Density (0), $X_{PS}$ (1), $Y_{PS}$ (2), Gaussian Flux Density (3), $X_G$ (4), $Y_G$ (5), Major Axis (6), where $X$ and $Y$ refer to the corresponding locations of the component in the Eastern and Northern directions from the phase centre respectively. 

\label{sec:app_vbc}

\subsection{Effect of $u-v$ coverage on correlations}

\begin{figure*}
    \centering
    \includegraphics[width=0.9\linewidth]{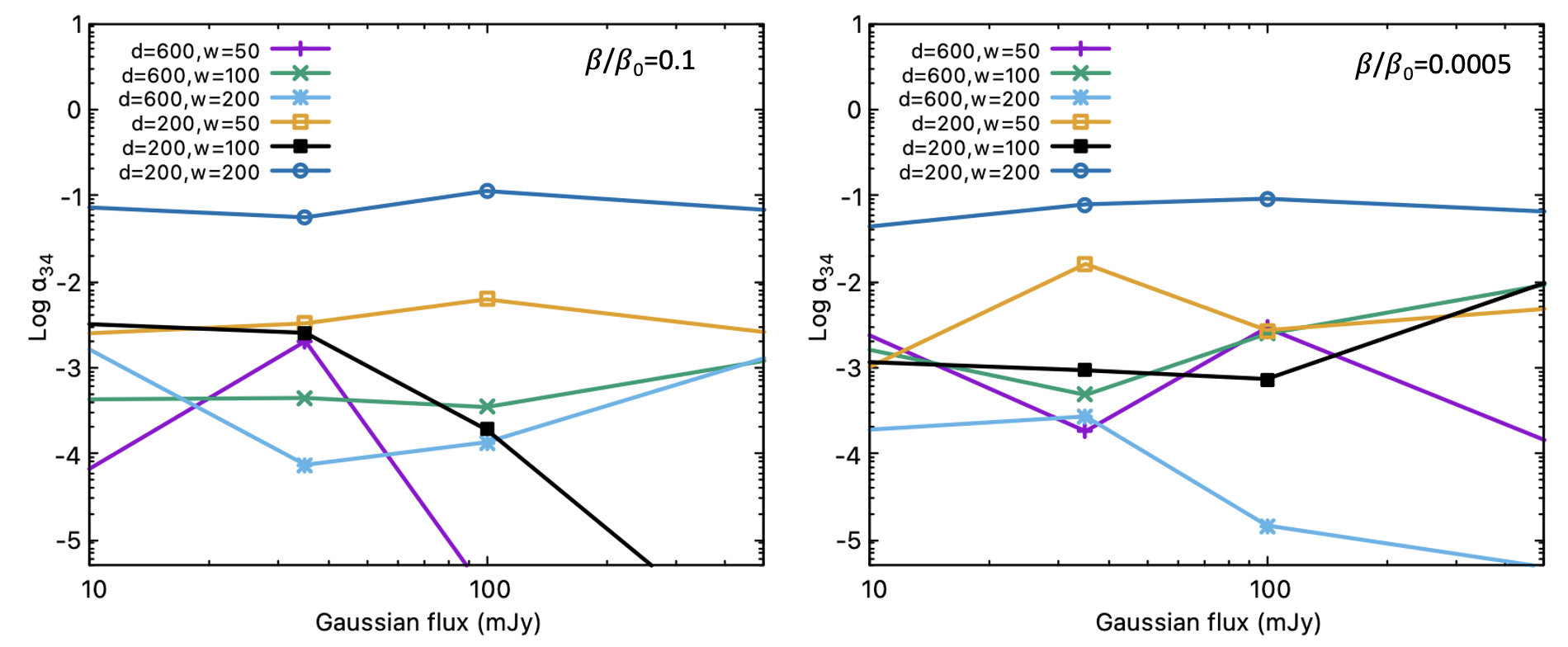}
    \caption{Figure shows $\alpha_{34}$, or the square of the correlation between the Gaussian flux density (3) and the Gaussian X position $X_G$ (3), in log-scale v/s Gaussian flux density for different parameter configurations, namely for $d=200, 600$ and major axis $w=50, 100, 200$ and for different $u-v$ coverage $\beta$. Left: $\alpha_{34}$ v/s Gaussian flux density for $\beta/\beta_0=0.1$. Right: $\alpha_{34}$ v/s Gaussian flux density for $\beta/\beta_0=0.0005$.} 
    \label{fig:corr34_uvonly}
\end{figure*}

Figure \ref{fig:corr34_uvonly} shows $\alpha_{34}$, or the square of the correlation between the Gaussian flux density (3) and the Gaussian X position $X_G$ (4) v/s Gaussian flux density for different parameter configurations and $\beta/\beta_0$. None of the panels show any significant patterns or correlations dependent on the Gaussian flux density and the graphs are mostly similar. Of special note is the specific case of $d=200, w=200$, which shows a hint of relatively stronger correlation for all flux values for both $\beta$. This can arise due to the fact that the boundary of the Gaussian coincides with the location of the point source, as $w=200$, making the flux more vulnerable to being mediated negatively by the core flux density, as in Figure \ref{fig:corr03_uvonly}. This implies that at farther distances away from the core (large $d$), fluctuations in both distance and flux would be positively correlated.

\begin{figure*}
    \centering
    \includegraphics[width=0.9\linewidth]{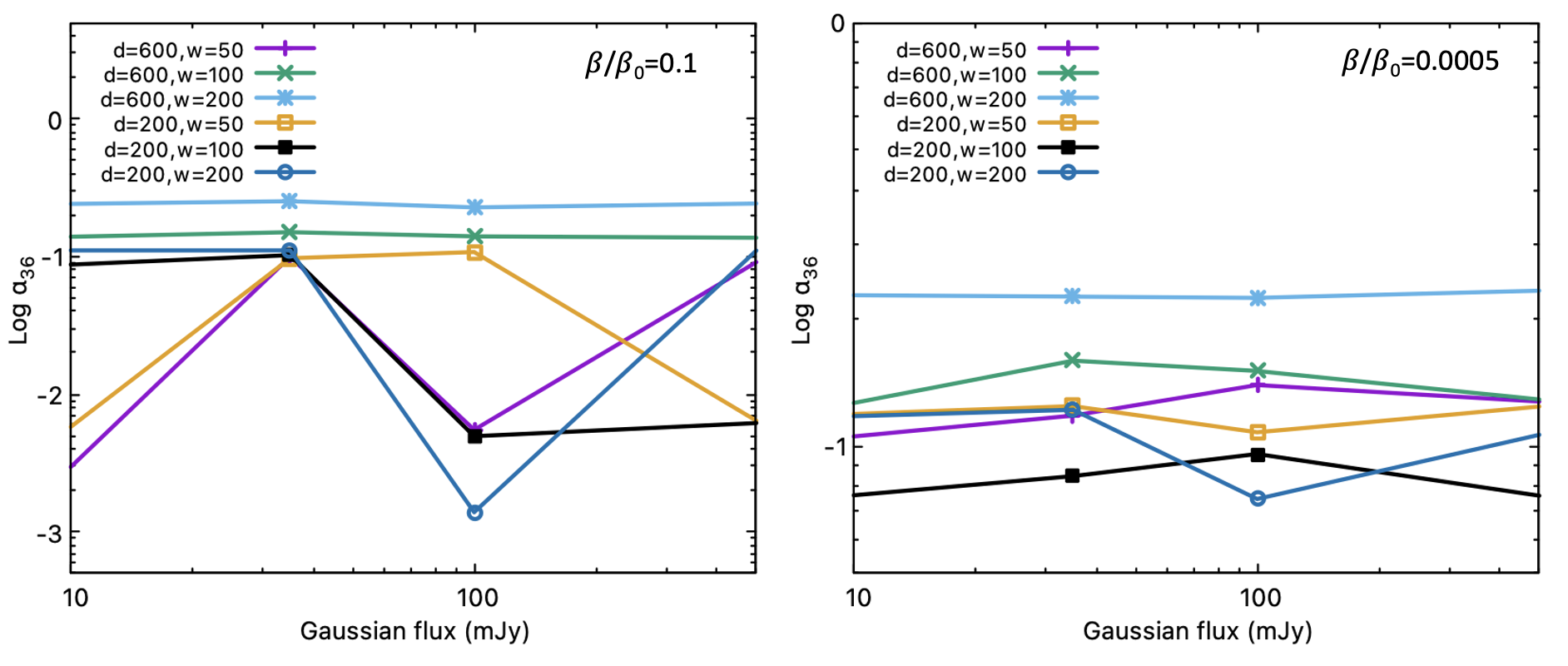}
    \caption{Figure shows $\alpha_{36}$, or the square of the correlation between the Gaussian flux density (3) and the Gaussian Major Axis $w$ (4), in log-scale v/s Gaussian flux density for different parameter configurations, namely for $d=200, 600$ and major axis $w=50, 100, 200$ and for different $u-v$ coverage $\beta$. Left: $\alpha_{36}$ v/s Gaussian flux density for $\beta/\beta_0=0.1$. $\alpha_{36}$ increases steadily on average with increase in flux. The correlation is particularly higher for the Gaussian farther away from the core ($d=600$) and of a larger size $w$. Right: $\alpha_{36}$ v/s Gaussian flux density for $\beta/\beta_0=0.0005$. $\alpha_{36}$ rises much more steadily than the left panel and all the samples reach a high correlation $\gtrsim0.05$ at the highest flux.}
    \label{fig:corr36_uvonly}
\end{figure*}

Figure \ref{fig:corr36_uvonly} shows $\alpha_{36}$, or the square of the correlation between the Gaussian flux density (3) and the Gaussian Major Axis $w$ (4) v/s Gaussian flux density for different parameter configurations and different $u-v$ coverage $\beta$. For the left panel, the correlation is significant through all Gaussian flux densities, but mainly for a larger distance from the core ($d=600$). The right panel has consistently large correlation for all parameter configurations, which seems to have been caused by the worsening of $u-v$ coverage. The correlation is particularly higher for the Gaussian farther away from the core ($d=600$) and of a larger size $w$. This trend of knots of larger sizes at a larger distance from the core having stronger correlation between flux and size can be explained as follows: to "conserve" visibility amplitude (which is not the same as flux), the real Gaussian width must increase with increase in flux density, resembling a positive correlation which we see here. For larger widths, this feature is more prominent. However, for distances close to the core, the Gaussian flux density more affected \textit{negatively} by the core flux as in Figure \ref{fig:corr03_uvonly}, thereby reducing its correlation with $w$. In addition, these correlations are more streamlined and strong at worse $u-v$ coverage. A related physical correlation is that of larger Gaussians imply larger flux densities; but since our sample is only from a model and not from a survey of jets, physical correlations are not very relevant.

\begin{figure*}
    \centering
    \includegraphics[width=0.9\linewidth]{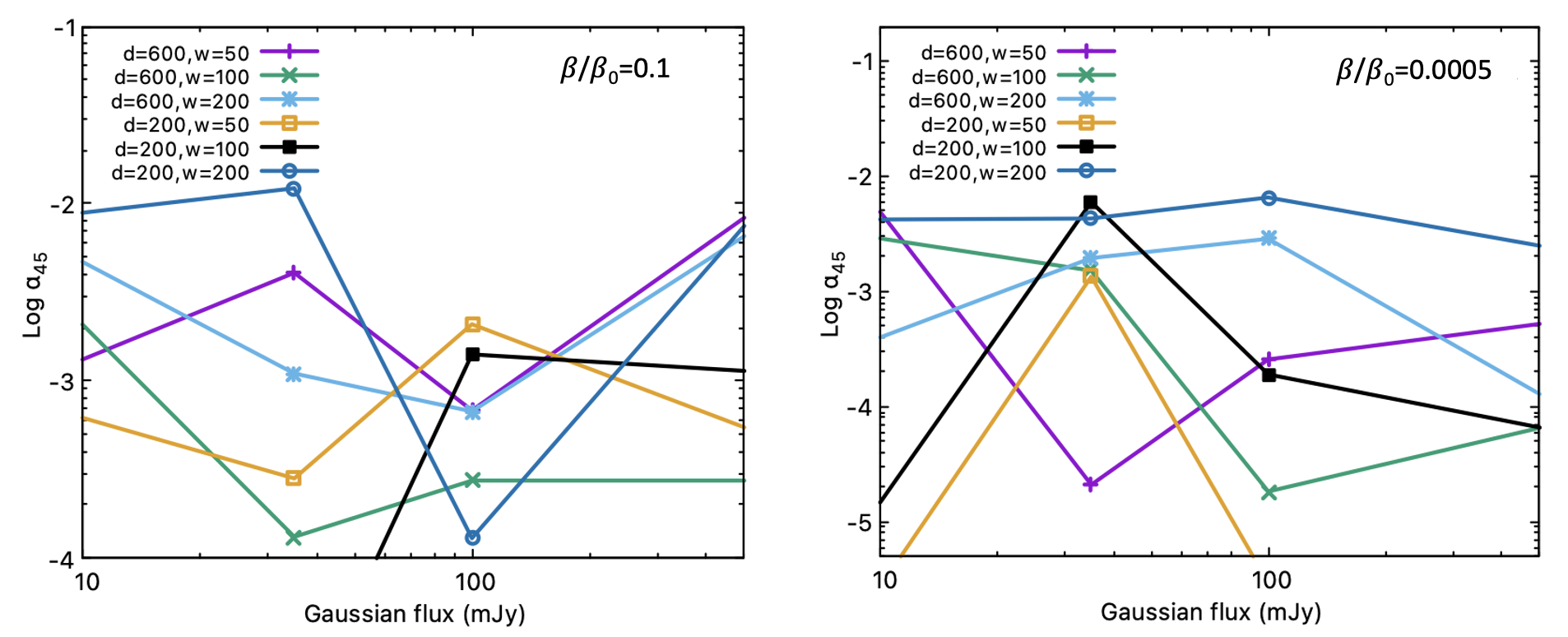}
    \caption{Figure shows correlation between Gaussian X ($X_G$) and Y positions ($Y_G$), or $\alpha_{46}$, in log-scale, v/s Gaussian flux density, for different parameters and different $\beta$. Left: $\alpha_{46}$ v/s Gaussian flux density for $\beta/\beta_0=0.1$. Right: $\alpha_{46}$ v/s Gaussian flux density for $\beta/\beta_0=0.0005$. None of the panels show significant dependence on Gaussian flux density or signs of correlation.}
    \label{fig:corr45_uvonly}
\end{figure*}

Figure \ref{fig:corr45_uvonly} shows correlation between Gaussian X ($X_G$) and Y positions ($Y_G$), or $\alpha_{45}$, v/s Gaussian flux density, for different parameters and different $\beta$. Other than the case of $w=200$, correlation is distinctly poor for both the panels for all flux densities and no specific pattern emerges. This is consistent with the fact that $X$ and $Y$ are fitting parameters rather than $d$, which would have otherwise resulted in a negative correlation between $X$ and $Y$.

\begin{figure*}
    \centering
    \includegraphics[width=0.9\linewidth]{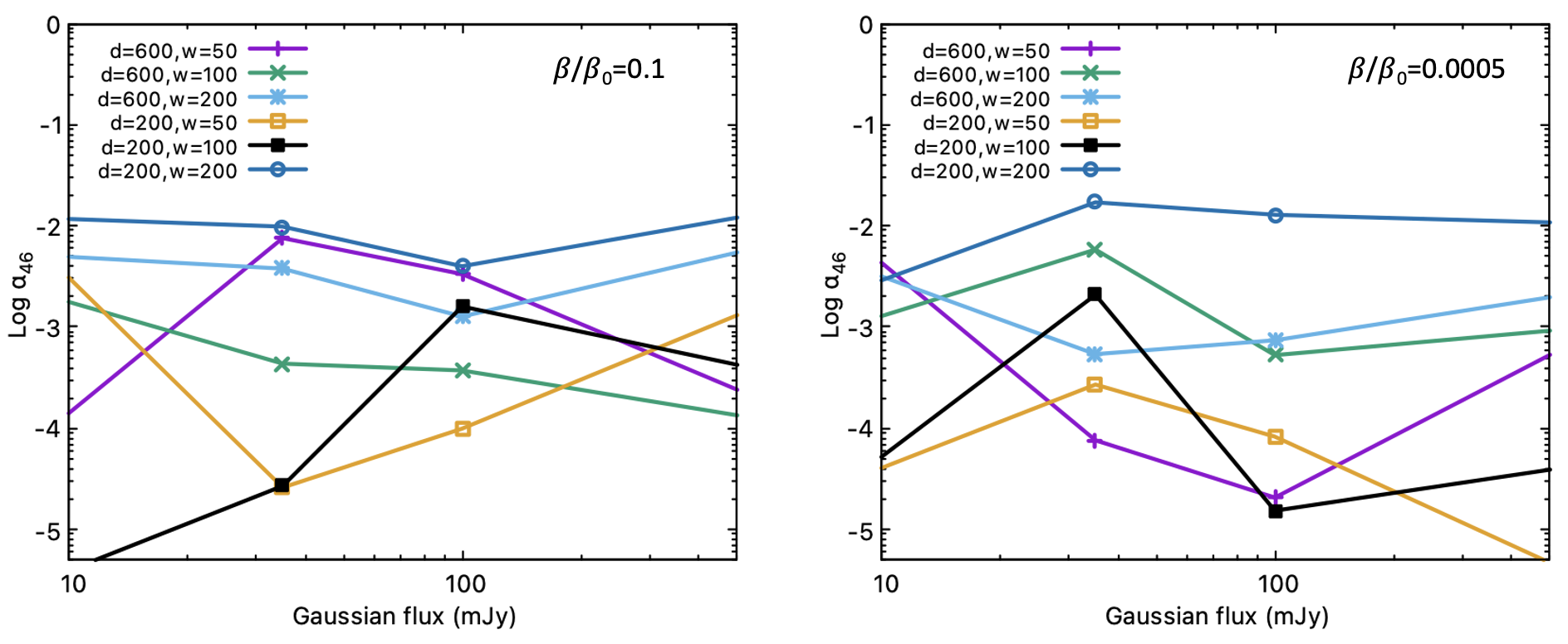}
    \caption{Figure shows $\alpha_{46}$, or the square of the correlation between $X_G$ and Major Axis $w$, in log-scale, v/s Gaussian flux density, for different parameters and different $\beta$. Left: $\alpha_{45}$ v/s Gaussian flux density for $\beta/\beta_0=0.1$. Right: $\alpha_{45}$ v/s Gaussian flux density for $\beta/\beta_0=0.0005$. No distinct correlation is observed.}
    \label{fig:corr46_uvonly}
\end{figure*}

Figure \ref{fig:corr46_uvonly} shows $\alpha_{46}$, or the square of the correlation between $X_G$ and Major Axis $w$ v/s Gaussian flux density, for different parameters and different $\beta$. No significant correlation or pattern emerges in any of the panels for any of the parameter configurations. However, the $d=200, w=200$ case shows relatively stronger (although low enough) correlation than all other pairs for both $\beta$. This is consistent with the fact that both position $X_G$ and flux are positively correlated for that specific case, in addition to the latter showing general positive correlation with $w$, as in Figures \ref{fig:corr34_uvonly} and \ref{fig:corr36_uvonly}.

\subsection{Effects of amplitude and phase errors on correlations}

\begin{figure*}
    \centering
    \includegraphics[width=\linewidth]{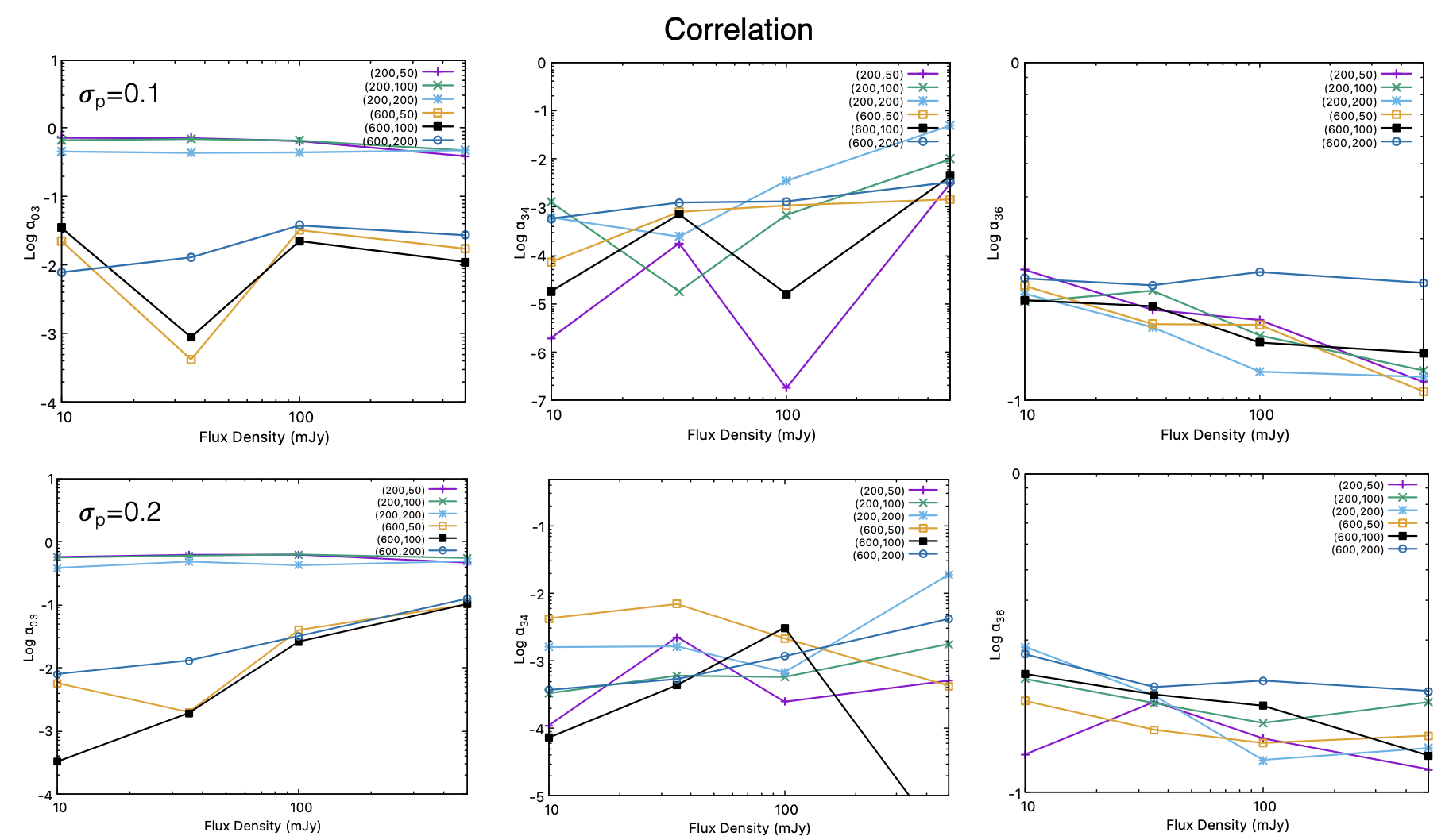}
    \caption{Figure shows the square of the correlations between the point source flux density (0) and the Gaussian flux density (3) ($\alpha_{03}$), Gaussian flux density (3) and Gaussian X position (4) ($\alpha_{34}$), Gaussian flux density (3) and Gaussian width (6) ($\alpha_{36}$) for different degrees of amplitude error $\sigma_g$ and phase error $\sigma_p$, as a function of the Gaussian flux. The legend shows the various configurations in a format (distance from the core, width of the Gaussian), or (d,w). Top panel: For 6 degree, or 0.1 radian, phase error. $\alpha_{03}$ and $\alpha_{36}$ are $\sim$ 1 dex larger than that in the absence of errors and are statistically significant, with the the largest through all the three cases dominated by the Gaussian configuration closer to the core and a larger width. $\alpha_{34}$, in contrast, show very weak correlations through almost all Gaussian flux densities.
    Bottom panel: For 12 degree, or 0.2 radian, phase error. For all the cases, the correlations for the $(200,w)$ case have mostly remained similar to the 0.1 radian case, while they have strengthened for the Gaussian at d=600 from the core, except for $\alpha_{36}$, which has hardly changed.}
    \label{fig:pe_corr}
\end{figure*}

In Figure \ref{fig:pe_corr}, for 6 degree, or 0.1 radian, phase error, $\alpha_{03}$ and $\alpha_{36}$ are $\sim$ 1 dex larger than that in the absence of errors and are statistically significant, with the the largest through all the three cases dominated by the Gaussian configuration closer to the core and a larger width. $\alpha_{34}$, in contrast, show very weak correlations through almost all Gaussian flux densities. In contrast for 12 degree, or 0.2 radian, phase error, the correlations for the $(200,w)$ case have mostly remained similar to the 0.1 radian case, while they have strengthened for the Gaussian at d=600 from the core, except for $\alpha_{36}$, which has hardly changed.

The two Figures \ref{fig:ge_corr} and \ref{fig:pe_corr} shine interesting light on the difference between the effects of amplitude and phase errors. The correlation between the point source flux density and the Gaussian flux ($\alpha_{03}$) density is, as expected to be larger for the case when the core is close to the "range of influence" of the Gaussian. This is similar for both the amplitude and phase error cases. However, the correlation between Gaussian flux density and X position ($\alpha_{34}$) is systematically smaller for the phase error case. It is a similar case for $\alpha_{36}$. This is expected since the amplitude error is a normalization error that mainly affects the parameters which encode the visibility amplitude, like source flux and size. In contrast, for a phase error, which must majorly affect the component positions, the correlations between Gaussian flux density, size and position must expectedly be lower.

\begin{figure*}
    \centering
    \includegraphics[width=\linewidth]{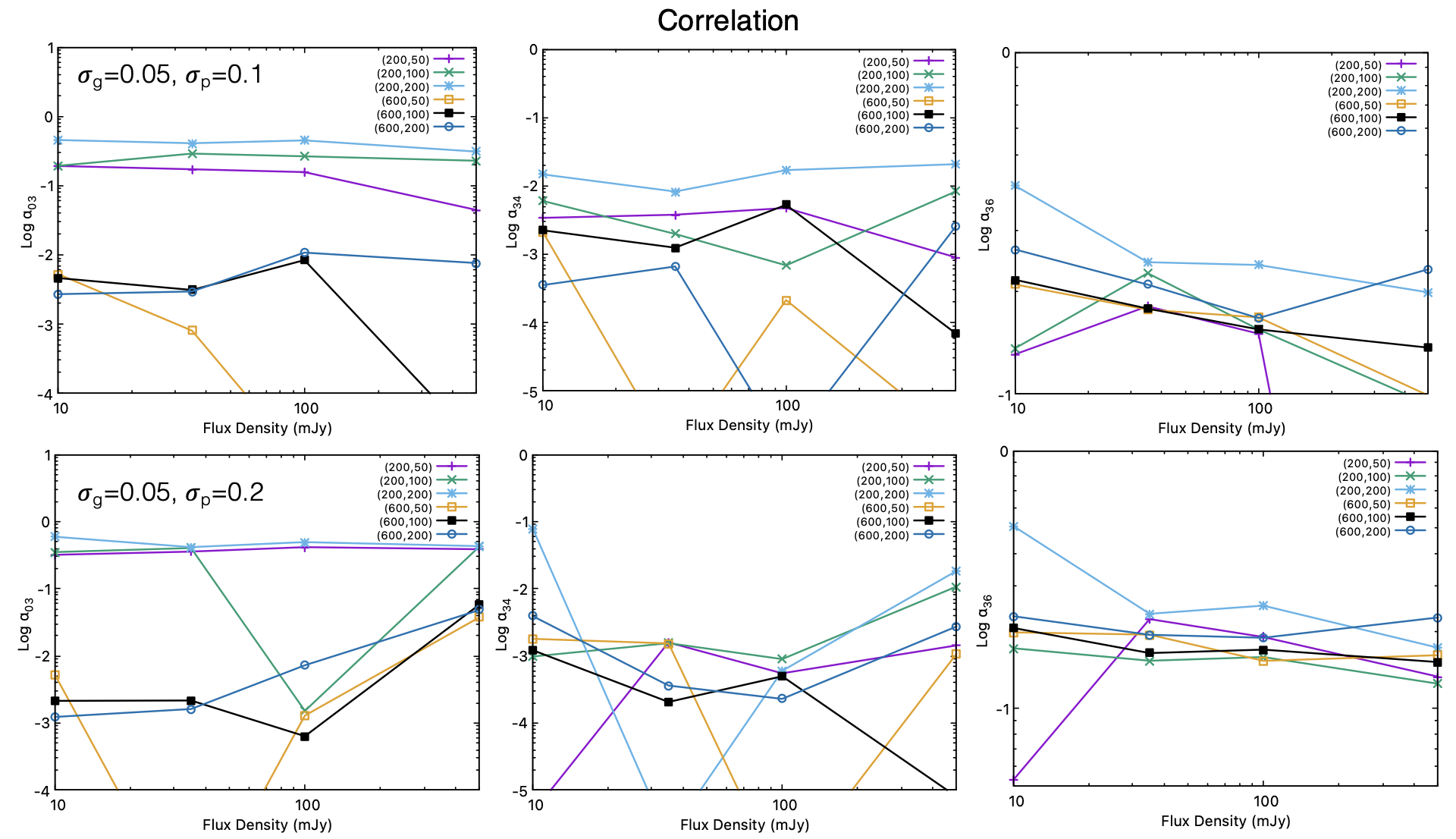}
    \caption{Figure shows the square of the correlations between the point source flux (0) and the Gaussian flux density (3) ($\alpha_{03}$), Gaussian flux density (3) and Gaussian X position (4) ($\alpha_{34}$), Gaussian flux density (3) and Gaussian width (6) ($\alpha_{36}$) for different degrees of amplitude error $\sigma_g$ and phase error $\sigma_p$, as a function of the Gaussian flux. The legend shows the various configurations in a format (distance from the core, width of the Gaussian), or (d,w). Top panel: For 5\% amplitude error and 0.1 radian (6 deg) phase error. $\alpha_{03}$ and $\alpha_{36}$ are large as both phase and amplitude errors individually showed similar behaviour. $\alpha_{34}$ is lower and clearly shows signs of "destroyed correlation" since amplitude errors create $\alpha_{34}$ correlations while phase errors destroy the same, like Figures \ref{fig:ge_corr} and \ref{fig:pe_corr}.
    Bottom panel: For 5\% amplitude error and 0.2 radian (12 deg) phase error. All are the similar to the top panel, except for $\alpha_{34}$ which further goes down on an average as the phase error increases.}
    \label{fig:gepe1_corr}
\end{figure*}

\begin{figure*}
    \centering
    \includegraphics[width=\linewidth]{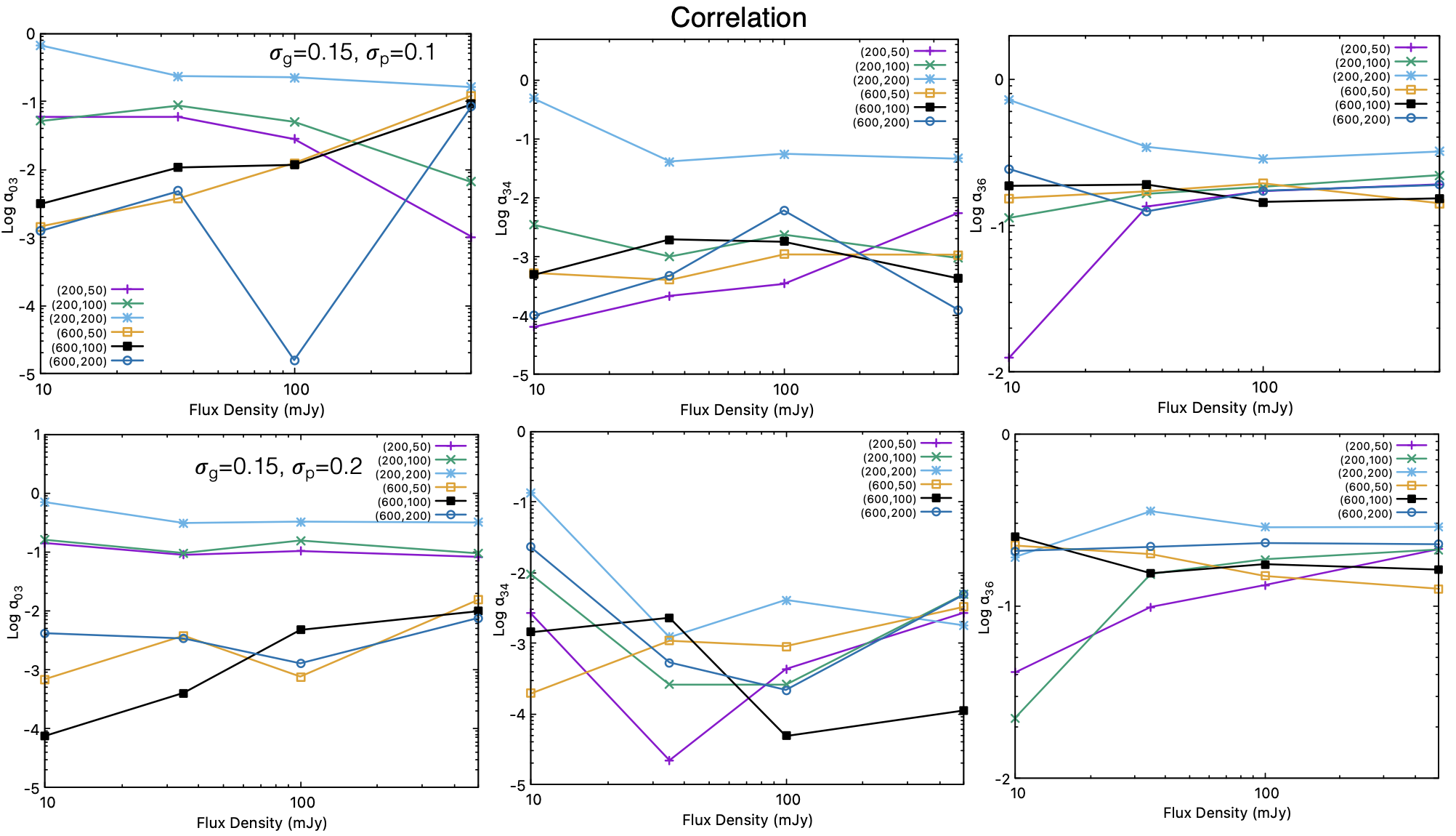}
    \caption{Figure shows the square of the correlations between the point source flux (0) and the Gaussian flux density (3) ($\alpha_{03}$), Gaussian flux density (3) and Gaussian X position (4) ($\alpha_{34}$), Gaussian flux density (3) and Gaussian width (6) ($\alpha_{36}$) for different degrees of amplitude error $\sigma_g$ and phase error $\sigma_p$, as a function of the Gaussian flux. The legend shows the various configurations in a format (distance from the core, width of the Gaussian), or (d,w). Top panel: For 15\% amplitude error and 0.1 radian (6 deg) phase error. $\alpha_{03}$ and $\alpha_{36}$ are large as both phase and amplitude errors individually showed similar behaviour, which is the same as Figure \ref{fig:gepe1_corr}. However, since the amplitude error is higher, $\alpha_{03}$ and $\alpha_{36}$ are higher than than Figure \ref{fig:gepe1_corr}. $\alpha_{34}$ is low, showing signs of "destroyed correlation" like Figures \ref{fig:ge_corr} and \ref{fig:pe_corr}. However, it is higher than Figure \ref{fig:gepe1_corr} since the amplitude error is higher, which has "restored" correlation.
    Bottom panel: For 15\% amplitude error and 0.2 radian (12 deg) phase error. All are the similar to the top panel, except for $\alpha_{34}$ which further goes down on an average as the phase error increases.}
    \label{fig:gepe2_corr}
\end{figure*}

Figures \ref{fig:gepe1_corr} and \ref{fig:gepe2_corr} show the total effect of both amplitude and phase errors in parameter correlations. Figure \ref{fig:gepe1_corr} shows $\alpha$ for 5\% amplitude error for each of 0.1 and 0.2 radian phase errors, while Figure \ref{fig:gepe2_corr} show the same but for 15\% amplitude error. As observed in Figures \ref{fig:ge_corr} and \ref{fig:pe_corr}, while $\alpha_{03}$ and $\alpha_{36}$ are systematically higher for both amplitude and phase errors, $\alpha_{34}$ is particularly low for phase errors, implying they destroy correlations between the Gaussian flux density and its X position, or essentially "not give rise" to correlations. Therefore $\alpha_{03}$ and $\alpha_{36}$ must be higher for all the combined cases of amplitude and phase errors, exactly as we observe. In contrast, $\alpha_{34}$ must be lower for the same amplitude error but higher phase error whereas higher for the same phase error but higher amplitude error, which is also what we 
observe in Figures \ref{fig:gepe1_corr} and \ref{fig:gepe2_corr}. Situations where both amplitude and phase errors occur are the most common and therefore these correlations need to be taken into account when fitting models to interferometric data. We have discussed this further in the Section 4.3.

\subsection{Effects of amplitude and phase errors on variances}

\begin{figure*}
    \centering
    \includegraphics[width=\linewidth]{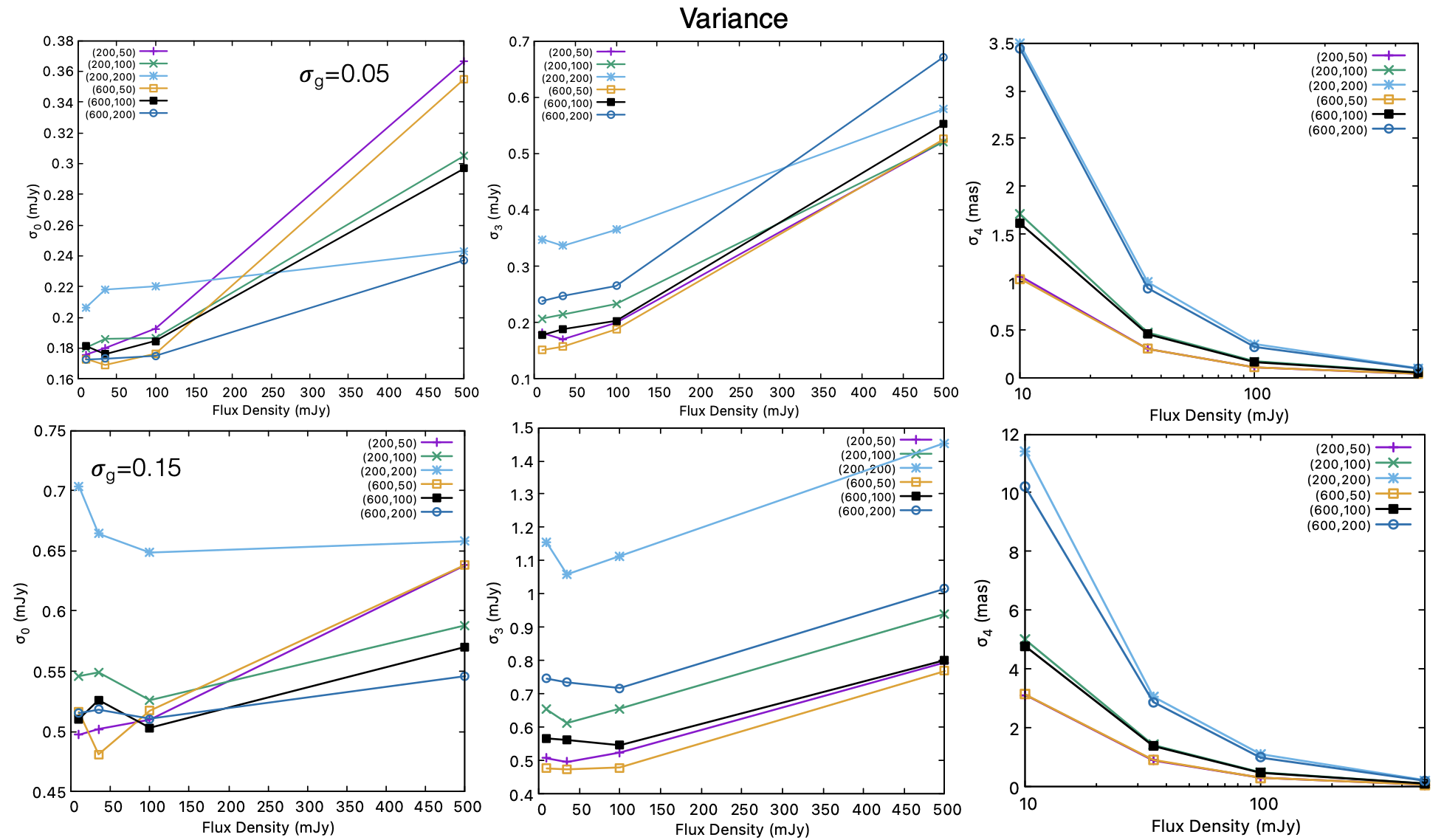}
    \caption{Figure shows the standard deviations of the determined best-fit point source flux (0), Gaussian flux density (3) and the Gaussian X position (4), for different degrees of amplitude error $\sigma_g$, as a function of the Gaussian flux. The legend shows the various configurations in a format (distance from the core, width of the Gaussian), or $(d,w)$. Top panel: For 5\% amplitude error. Both $\sigma_0$ and $\sigma_3$ increase with increase in Gaussian flux density, implying a constant fractional variance. While $\sigma_4$ decreases as expected with increase in Gaussian flux density, it is also lowest for the most compact component. Bottom panel: For 15\% amplitude error. The behaviour is very broadly similar to the top panel, except the uncertainties are 3-4 times larger. The faintest Gaussian closest to the point source and largest in size shows higher uncertainties on average, as it comes "under the influence" of the point source.}
    \label{fig:ge_sig}
\end{figure*}

In Figure \ref{fig:ge_sig}, for both cases of amplitude error, $\sigma_0$ and $\sigma_3$ steadily increase (almost linearly) with increase in Gaussian flux. While it implies that the Gaussian flux densities have a conserved fractional variance with increase in the same, the behaviour is also similar for the core flux. The latter is expected from a "competition" between the component flux densities as the Gaussian brightens. On an average, $\alpha_0$ and $\alpha_3$ are higher for the faintest Gaussian component and that which comes under the influence of the core emission. This is majorly intuitively expected. $\sigma$ increases by 3-4 times for both the PS and the Gaussian flux densities as the amplitude error is tripled from 5\% to 15\%. The behaviour of the Gaussian X position variance is novel. For both the cases of amplitude error, while it goes down with increase in Gaussian flux density as expected, $\sigma$ goes to a minimum when the Gaussian is most compact. This implies that more compact and brighter jet "knots" can be localized better, in accordance with our intuition. Furthermore, $\sigma$ almost quadruples from 5 to 15\% amplitude error.

\begin{figure*}
    \centering
    \includegraphics[width=\linewidth]{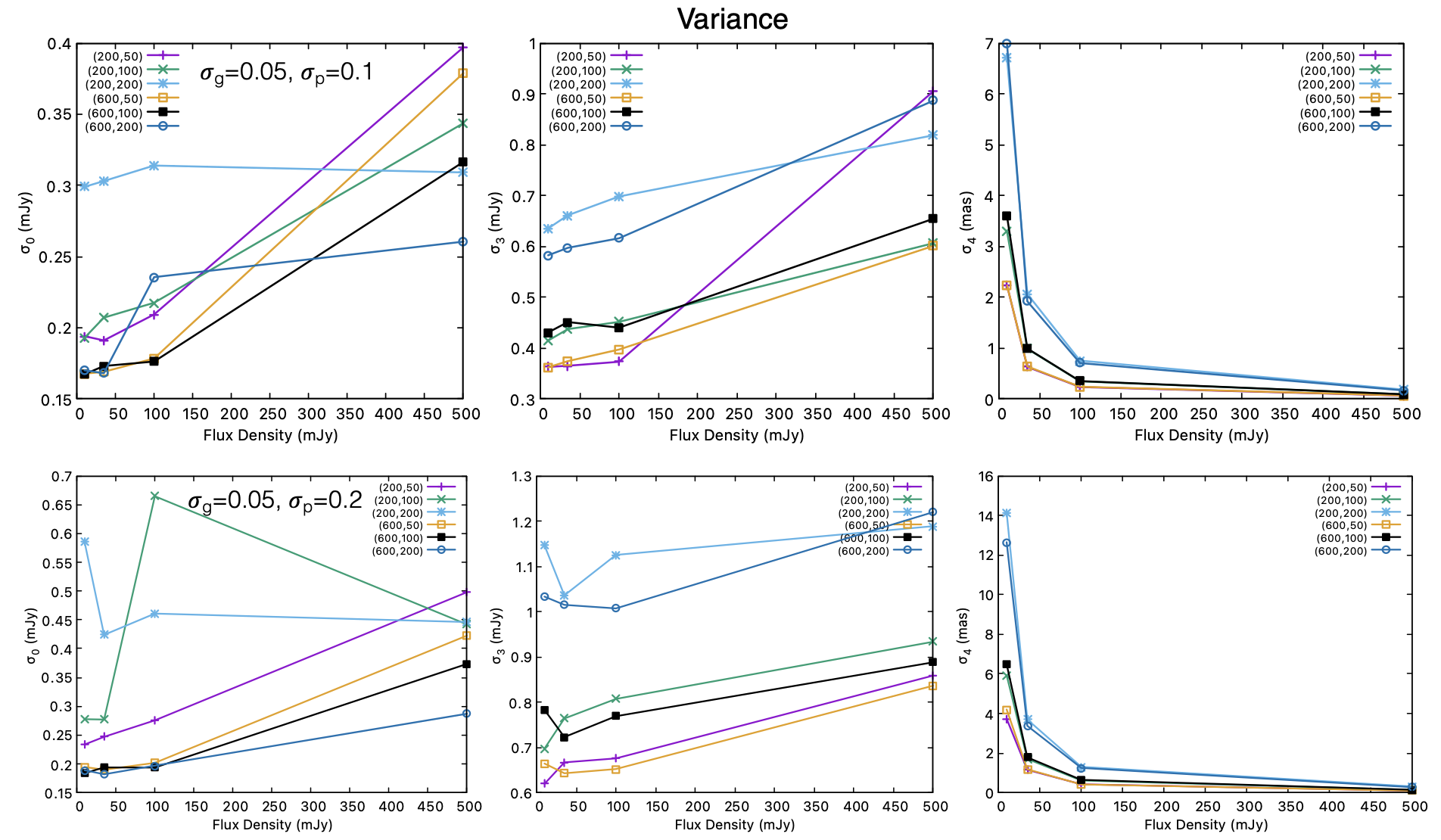}
    \caption{Figure shows the standard deviations of the determined best-fit point source flux (0), Gaussian flux density (3) and the Gaussian X position (4), for different degrees of amplitude error $\sigma_g$ and phase error $\sigma_p$, as a function of the Gaussian flux. The legend shows the various configurations in a format (distance from the core, width of the Gaussian), or $(d,w)$. Top panel: For 5\% amplitude and 0.1 radian phase error. Both $\sigma_0$ and $\sigma_3$ increase with increase in Gaussian flux density on average, implying a constant fractional variance. $\sigma_0$ is lower when the Gaussian is away from the core while $\sigma_3$ is lower when the Gaussian is more compact, as expected. $\sigma_4$ is considerably higher than the previous discrete cases. While $\sigma_4$ decreases as expected with increase in Gaussian flux density, it is also lowest for the most compact component. Bottom panel: For 5\% amplitude and 0.2 radian phase error. The behaviour of $\sigma_0$ and $\sigma_3$ is very broadly similar to the top panel, with similar uncertainties. $\sigma_4$ is larger than the top panel as expected, and reaches large $>10$ mas values for fainter and more extended Gaussians.}
    \label{fig:gepe1_sig}
\end{figure*}

\begin{figure*}
    \centering
    \includegraphics[width=\linewidth]{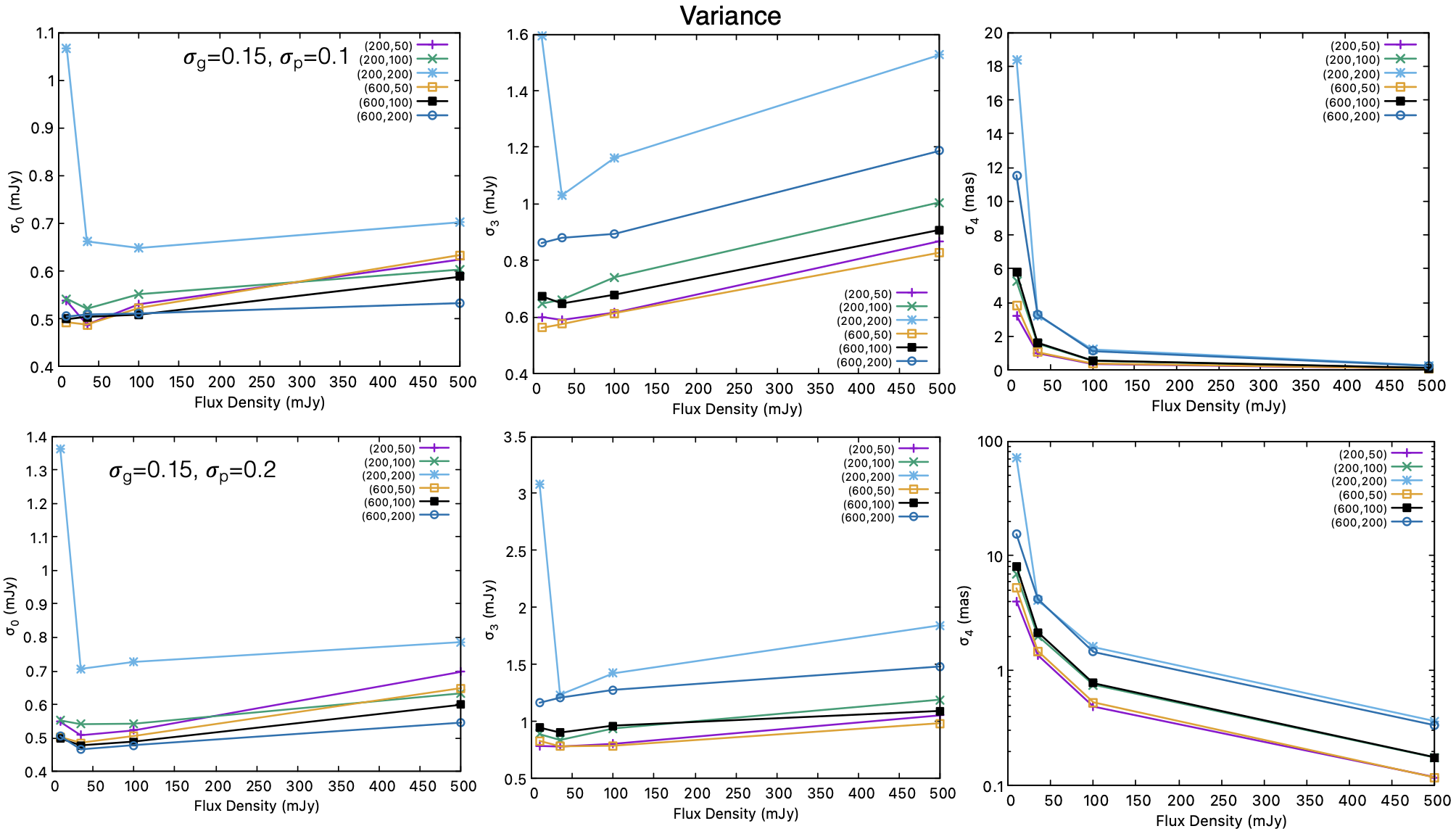}
    \caption{Figure shows the standard deviations of the determined best-fit point source flux (0), Gaussian flux density (3) and the Gaussian X position (4), for different degrees of amplitude error $\sigma_g$ and phase error $\sigma_p$, as a function of the Gaussian flux. The legend shows the various configurations in a format (distance from the core, width of the Gaussian), or $(d,w)$. Top panel: For 15\% amplitude and 0.1 radian phase error. The behaviours of $\sigma_0$, $\sigma_3$ and $\sigma_4$ are roughly similar to Figure \ref{fig:gepe1_sig} except larger in magnitude. Bottom panel: For 15\% amplitude and 0.2 radian phase error. $\sigma_0$, $\sigma_3$ and $\sigma_4$ follow similar patterns as top panel, but they are larger in magnitude than all previous figures. $\sigma_4$ reaches values close to 100 mas for the most extended and faint component.}
    \label{fig:gepe2_sig}
\end{figure*}

Figures \ref{fig:gepe1_sig} and \ref{fig:gepe2_sig} show the combined effects of amplitude and phase errors on the parameter variance, which can be intuitively understood by considering the separate effects of amplitude and phase errors in Figures \ref{fig:ge_sig} and \ref{fig:pe_sig}. For example, when the amplitude error is constant, change in phase error must follow that in Figure \ref{fig:pe_sig}, and vice versa as in Figure \ref{fig:ge_sig}. In Figure \ref{fig:gepe1_sig}, for 5\% amplitude and both the cases of 0.1 and 0.2 radian phase error, $\sigma_0$ and $\sigma_3$ increase with Gaussian flux density on an average, implying a pseduo constant fractional variance. $\sigma_0$ is minimum when the Gaussian is away from the influence of the core. $\sigma_3$ is minimum expectedly for the more compact Gaussian case. There is a similar trend in Figure \ref{fig:gepe2_sig} for $\sigma_0$ and $\sigma_3$, with the magnitude few times higher since the amplitude error is higher. The case of $\sigma_4$ across the two figures is enlightening. $\sigma_4$ reaches $>5$ mas for the fainter Gaussians for 0.1 radian phase errors, while hovers between $10-100$ mas for 0.2 radian phase errors. For higher Gaussian flux densities it drops rapidly. This implies that in real datasets this must be considered if the Gaussian is much fainter than the core.

\subsection{Effects of amplitude and phase errors on biases}

\begin{figure*}
    \centering
    \includegraphics[width=\linewidth]{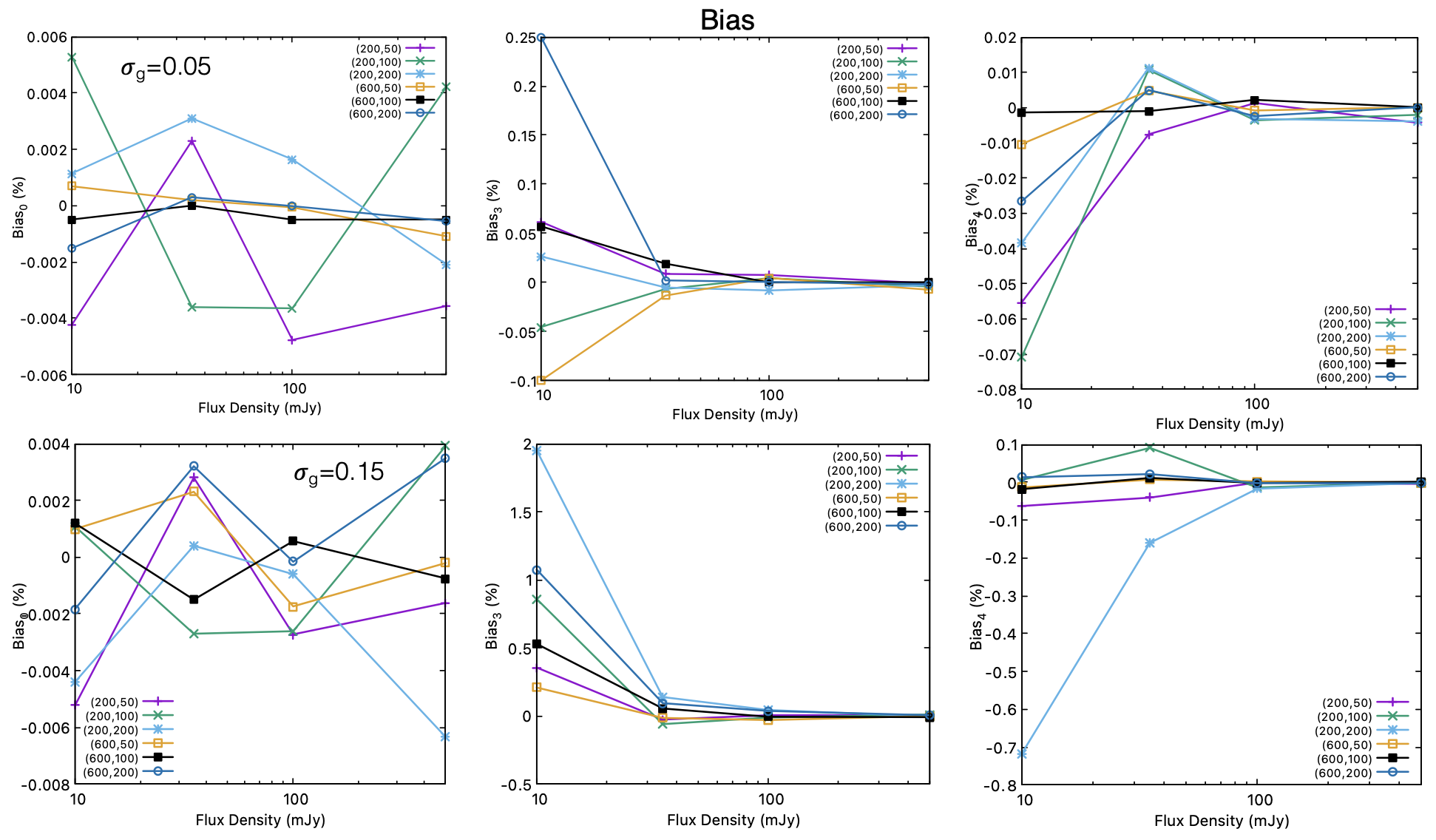}
    \caption{Figure shows the biases of the determined best-fit point source flux (0), Gaussian flux density (3) and the Gaussian X position (4), for different degrees of amplitude error $\sigma_g$, as a function of the Gaussian flux. The legend shows the various configurations in a format (distance from the core, width of the Gaussian), or $(d,w)$. Top panel: For 5\% amplitude error. $Bias_0$ is expectedly very close to zero. $Bias_3$ is mildly positive and $Bias_4$ is mildly negative, both more so for the components closer to the core and larger in size. The former is expected as the Gaussian "competes" with the core during the flux fitting. The latter is also expected as then the position of Gaussian would be biased towards the core, or essentially negatively biased. Both $Bias_3$ and $Bias_4$ reach zero in the limit of large flux. Bottom panel: For 15\% amplitude error. All the biases are very similar in shape to the top panel. While $Bias_0$ has not changed, $Bias_3$ and $Bias_4$ have increased by an order of magnitude.}
    \label{fig:ge_bias}
\end{figure*}

The bias for a parameter $x$ is simply given as $(x_{\rm obs}-x_{\rm expected})/x_{\rm expected}$. In Figure \ref{fig:ge_bias}, for 5\% and 15\% amplitude error we see roughly similar behaviour for biases for parameter indices 0, 3 and 4. The point source flux density expectedly not biased since it is very localized and is the brightest component, while the Gaussian flux density (3) is more biased towards positive flux densities for smaller values of $d$ and larger values of $w$. This is expected since then it is more under the influence of the core and it tends to "take away" some of the core flux since the latter is much larger. The bias in the Gaussian X position is negative for smaller $d$ and larger $w$, for very similar reasons. The biases increase by an order of magnitude to $\sim$ 1\% and -0.5\% for parameters (3) and (4) respectively for 15\% amplitude error, while remain the same for the point source flux. This is partly expected since larger amplitude errors further "blur" the line of difference between the Gaussian and the core in the $u-v$ plane. Also as expected, the biases $Bias_3$ and $Bias_4$ reach zero for larger Gaussian flux densities, where the convergence is slower for larger amplitude error.

\begin{figure*}
    \centering
    \includegraphics[width=\linewidth]{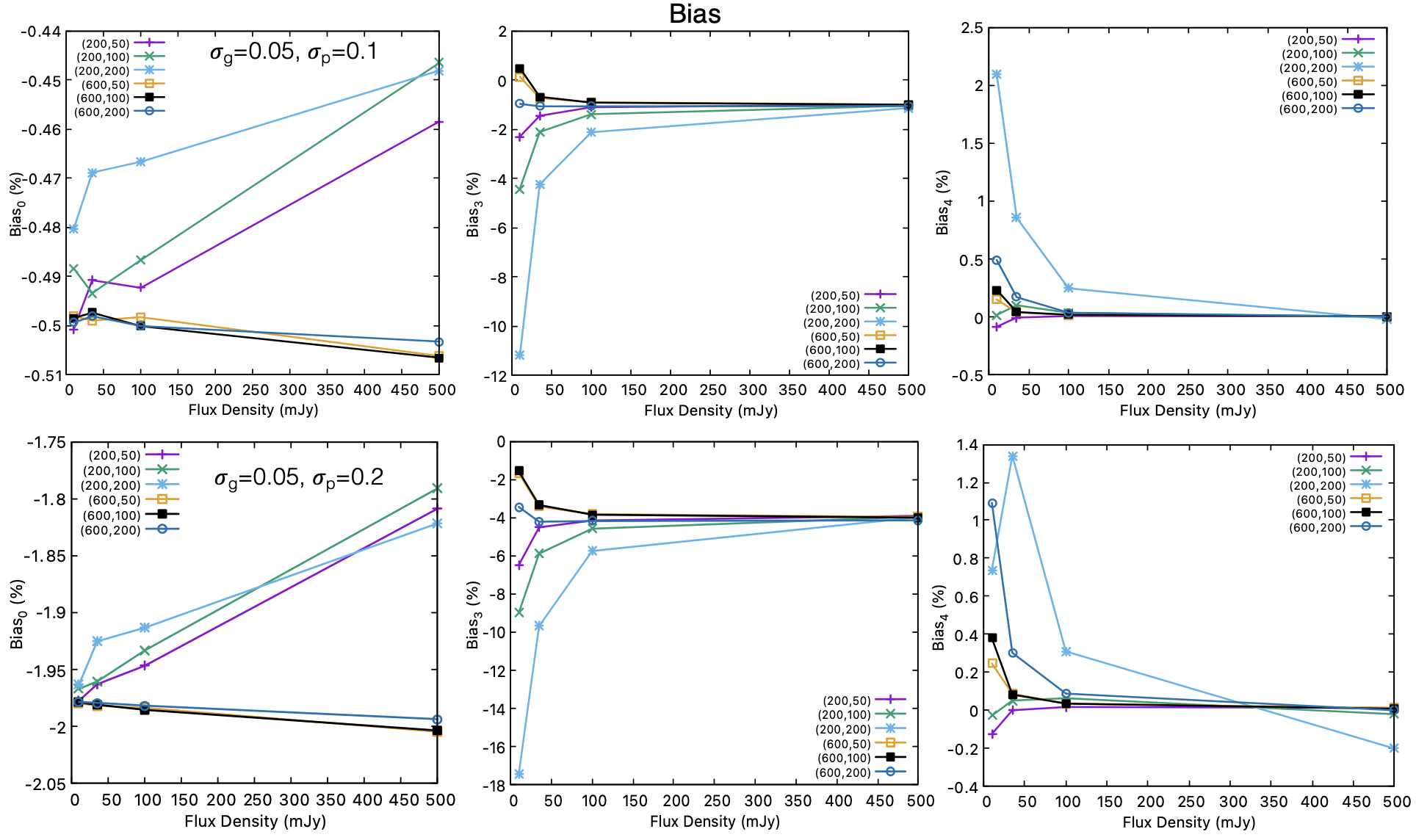}
    \caption{Figure shows the biases of the determined best-fit point source flux (0), Gaussian flux density (3) and the Gaussian X position (4), for different degrees of amplitude error $\sigma_g$ and phase error $\sigma_p$, as a function of the Gaussian flux. The legend shows the various configurations in a format (distance from the core, width of the Gaussian), or $(d,w)$. Top panel: For 5\% amplitude error and 0.1 radian phase error. $Bias_0$ and $Bias_3$ are negative with the distinct pattern observed in Figure \ref{fig:pe_bias}, implying dominance of phase errors over amplitude errors. $Bias_4$ is positive and goes to zero at large flux density. Bottom Panel: Similar behaviour for $Bias_0$ and $Bias_3$ compared to top panel, but are only 3-4 times more negative. $Bias_4$ is similarly behaved with similar bias strengths.}
    \label{fig:gepe1_bias}
\end{figure*}

\begin{figure*}
    \centering
    \includegraphics[width=\linewidth]{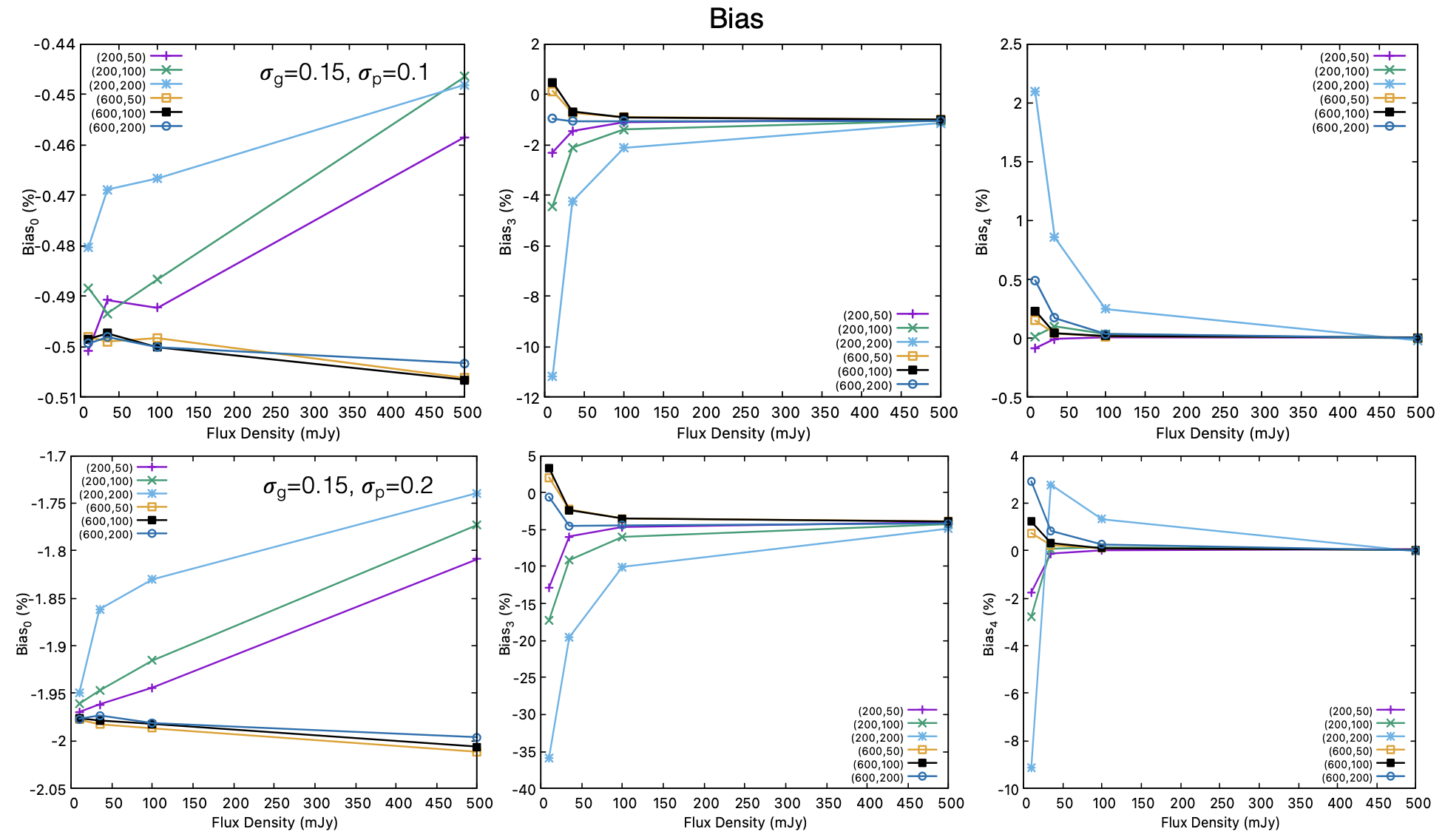}
    \caption{Figure shows the biases of the determined best-fit point source flux (0), Gaussian flux density (3) and the Gaussian X position (4), for different degrees of phase error $\sigma_p$, as a function of the Gaussian flux. The legend shows the various configurations in a format (distance from the core, width of the Gaussian), or $(d,w)$. Top panel: For 5\% amplitude error and 0.1 radian phase error. $Bias_0$ is negative with the similar pattern as well as the similar magnitude observed in Figure \ref{fig:gepe1_bias}. $Bias_3$, in contrast, although follows a similar pattern, is 3-4 times more negative than Figure \ref{fig:gepe1_bias}, implying an increase in amplitude error has worsened the bias. $Bias_4$ is positive and goes to zero at large flux density. Bottom Panel: Similar behaviour for $Bias_0$ and $Bias_3$ compared to top panel, but are only 3-4 times more negative. $Bias_4$ is negative for most components, unlike top panel and Figure \ref{fig:gepe1_bias}.}
    \label{fig:gepe2_bias}
\end{figure*}

Figures \ref{fig:gepe1_bias} and \ref{fig:gepe2_bias} show the combined effects of amplitude and phase errors on the parameter bias, which can also be intuitively understood by considering the separate effects of amplitude and phase errors in Figures \ref{fig:ge_bias} and \ref{fig:pe_bias}. In Figure \ref{fig:gepe1_bias}, for 5\% amplitude and both the cases of 0.1 and 0.2 radian phase error, $Bias_0$ and $Bias_3$ are strongly negative, implying dominance of phase error. The bias is more negative for the higher phase error, as expected. The same picture carries forward to Figure \ref{fig:gepe2_bias}, where the amplitude error is 15\%. In this case the bias is similar for the point source flux density, but more negative for the Gaussian flux density ($Bias_3$), where the amplitude error must have exacerbated the negative bias. However, $Bias_4$ has no strict pattern across both the figures for different values of amplitude and phase errors. It is mostly positive (around few \%) except for $\sigma_g=0.15$ and $\sigma_p=0.2$ configuration, where it is strongly negative (close to 5-10\%) for the Gaussian closer to the core. The absence of any specific pattern hints at a non-trivial effect of phase and amplitude errors, which depends on the specific values of the errors chosen and the source structure. Therefore, this can only be determined on a case by case basis, without any generalization. However, $Bias_4$, in \textit{all} the figures, tends to 0 in the limit of large Gaussian flux; hence such a bias can be "fixed" if the component is bright enough, unlike $Bias_3$ which retains a constant bias even at large flux densities.

\bibliography{version1,version1_second}{}
\bibliographystyle{mnras}

\end{document}